\documentclass[twocolumn]{aastex62}

\usepackage{graphicx}
\usepackage{amsmath}
\usepackage{appendix}

\graphicspath{{./}{figures/}}

\received{November 10, 2019}
\revised{December 26, 2019}
\accepted{December 29, 2019}
\submitjournal{ApJS}

\shorttitle{Telluric-Corrected Solar Atlas}
\shortauthors{Baker et al.}

\begin{document}

\title{The IAG Solar Flux Atlas: Telluric Correction With a Semi-Empirical Model}

\author[0000-0002-0786-7307]{Ashley D. Baker}
\affiliation{University of Pennsylvania 
Department of Physics and Astronomy,  
209 S 33rd St, 
Philadelphia, PA 19104, USA}
\author{Cullen H. Blake}
\affiliation{University of Pennsylvania 
Department of Physics and Astronomy,  
209 S 33rd St, 
Philadelphia, PA 19104, USA}
\author{Ansgar Reiners}
\affiliation{Georg-August Universit\"{a}t G\"{o}ttingen, Institut f\"{u}r Astrophysik, Friedrich-Hund-Platz 1, 37077 G\"{o}ttingen, Germany}

\correspondingauthor{Ashley Baker}
\email{ashbaker@sas.upenn.edu}

\begin{abstract}
Observations of the Sun as a star have been key to guiding models of stellar atmospheres and additionally provide useful insights on the effects of granulation and stellar activity on radial velocity measurements. Most high resolution solar atlases contain telluric lines that span the optical and limit the spectral regions useful for analysis. We present here a telluric-corrected solar atlas covering 0.5-1.0~$\mu$m derived from solar spectra taken with a Fourier Transform Spectrograph (FTS) at the Institut f\"{u}r Astrophysik, G\"{o}ttingen. This atlas is the highest resolution spectrum with a wavelength calibration precise to $\pm$10~m~s$^{-1}$ across this 500~nm spectral window. We find that the atlas matches to within 3\% of the telluric-corrected Kitt Peak atlas in regions containing telluric absorption weaker than 50\% in transmission. The telluric component of the spectral data is fit with a semi-empirical model composed of Lorentz profiles initialized to the HITRAN parameters for each absorption feature. Comparisons between the best-fit telluric parameters describing the Lorentz profile for each absorption feature to the original HITRAN values in general show excellent agreement considering the effects atmospheric pressure and temperature have on our final parameters. However, we identify a small subset of absorption features with larger offsets relative to the catalogued line parameters. We make our final solar atlas available online. We additionally make available the telluric spectra extracted from the data that, given the high resolution of the spectrum, would be useful for studying the time evolution of telluric line shapes and their impact on Doppler measurements.
\end{abstract}


\section{Introduction} \label{sec:intro}
The Sun is the most well-studied star and serves as a benchmark for our understanding of stellar physics. High resolution spectra of the Sun have been essential references for a variety of stellar and planetary studies that strive to understand atomic physics processes in the solar atmosphere \citep{intro_model,sun_aluminum}, determine the chemical abundances of other stars \citep{intro_asp09,intro_bru12}, and measure the radial velocities of solar system objects reflecting solar light \citep{intro_bon14}. More recently, the use of solar spectra has been helpful to understand the effects of magnetic activity and gravitational blueshift on radial velocity (RV) measurements \citep{harps_rv,Reiners16,harps15, harps_fit_challenge}. These studies strive to develop ways to reduce the effects of activity on stellar spectra that can limit the achieved RV measurement precision.

An additional application of solar spectral observations is the study of telluric lines themselves since the Sun serves as a bright back-light to the atmosphere that provides enough light for high signal to noise, high resolution telluric measurements. This enables the study of weak telluric lines, or micro-tellurics, that are of additional pertinence to exoplanet radial velocity studies due to their ability to impact precision RV measurements \citep{cunha14,sam_rv_err_budget,micro_telluric_peter,artigau14,telluric_budget}. Solar spectra have also proven useful to measure abundances of atmospheric gases \citep{toon_balloon,solar_occult,solar_o3} and to validate line parameter databases \citep{toon16}.

High resolution, disk integrated spectra of the Sun are difficult to obtain from space, but are crucial for studies that need to view the Sun as a star. Ground-based solar atlases generated with spectrographs having resolution too low to resolve the solar lines are useful, but suffer from the effects of an instrument-specific line spread profile (LSP). The convolution of the full observed spectrum with the LSP makes telluric removal challenging since a correction by dividing a telluric model is no longer mathematically exact. Furthermore, the convolution with the LSP complicates comparisons between data and models at different resolutions.

Several ground-based instruments have been successful at observing disk integrated spectra of the Sun at a resolution that fully resolves the solar lines. These include \cite{Kurucz84} and \cite{Reiners16}, which both utilized Fourier Transform Spectrographs (FTS). In a comparison of the wavelength solutions of these two atlases, \cite{Reiners16} showed that errors in the wavelength calibration of the Kitt Peak solar atlas were \textgreater 50~m s$^{-1}$ in regions blue of 473~nm and 20~m~s$^{-1}$ red of 850~nm whereas the Institut f\"{u}r Astrophysik, G\"{o}ttingen (IAG) solar flux atlas shows good agreement to within 10~m s$^{-1}$ with a HARPS laser frequency comb calibrated atlas \citep{Molaro13}, which was taken at slightly lower resolution covering a 100~nm range around 530~nm. Comparisons of the IAG atlas with a second Kitt Peak atlas derived slightly differently from the first (see \citealt{Wallace11}) shows even larger offsets.

These high resolution, disk integrated solar atlases are commonly used as a comparison to solar models that are important for studying non-local thermal equilibrium effects that change the line shapes of various molecular features, such as that of calcium \citep{calcium_line_tests}. Additionally, observing the Sun as a star and measuring the line bisectors provides important information for exoplanet radial velocity studies that must understand the limiting effects of granulation and star spots on their radial velocity measurements. For these applications, the Sun serves as a useful test case since high resolution and high signal-to-noise measurements can be achieved, which are necessary to compare measurements to models and study the effects of degrading the spectral resolution, as will commonly be the case for measurements of other stars \citep{cegla19, bisector_resolution}. Additionally, being able to image the stellar surface provides extra information in studying the effect of star spots. For all these applications, telluric lines can skew the measurements \citep{lars18_bisector} and limit the spectral regions that are useful for these studies \citep{sun_aluminum,sun_balloon}. 

Few efforts have been made to correct solar spectra for telluric lines, possibly because this is more difficult for the Sun due to the lack of telluric reference stars and the low Doppler shifts of the solar lines, meaning telluric lines do not shift significantly with respect to the solar features. In work by \cite{Kurucz06}, the Kitt Peak Solar Atlas (KPSA) was telluric-corrected using a full radiative transfer atmospheric model. Residuals were replaced by hand with lines connecting the boundaries of contaminated regions. In 2011 another disk-integrated solar atlas was observed at Kitt Peak by \cite{Wallace11} who corrected the spectrum for atmospheric absorption by using telluric data derived from disk-centered solar spectra. The improved wavelength calibration of the IAG atlas in addition to the lack of uncertainties on the Kitt Peak atlases motivates the derivation of a new telluric-corrected solar atlas derived from IAG solar spectral data. Here, we generate this telluric-corrected IAG solar flux atlas that has estimated uncertainties that capture the success of the telluric removal process and therefore makes it useful for studies that wish to mask or properly weight telluric-contaminated spectral regions. To achieve this, we develop a unique semi-empirical telluric fitting method that works well despite the small Doppler shifts of the solar lines that makes it challenging to dissociate them from overlapping telluric features.

In \S \ref{sec:data} we describe the data set used to generate this atlas and the pre-processing steps performed to determine the wavelength calibration and solar radial velocity for each spectrum. In \S \ref{sec:model} we describe the model framework and in \S \ref{sec:fit} we describe the fitting sequence and how we use the best-fit models to generate the output data products that include the final atlas and an archive of telluric spectra. We provide an analysis in \S \ref{sec:analyze} in which we compare our solar atlas to the KPSA and discuss our telluric model and findings related to the telluric line shape parameters. Finally, we conclude in \S \ref{sec:conclude}.

\section{Data}\label{sec:data}
The data used here for generating a telluric-free, disk-integrated solar spectrum were taken with the Vacuum Vertical Telescope\footnote{https://www.uni-goettingen.de/en/217813.html} (VVT) at the Institut f\"{u}r Astrophysik in G\"{o}ttingen, Germany. A siderostat mounted on the telescope directs light into an optical fiber that passes through an iodine cell before feeding the Fourier transform spectrograph \citep{lemke16}. The iodine cell serves to provide a wavelength calibration that is more accurate than the internal calibration of the FTS. We utilize a subset of spectra from a 20-day data set taken over the span of a year with each day having disk-integrated solar spectra recorded over a multiple hour span. Additionally, 450 spectra were taken using a halogen light source with the iodine cell in order to generate an iodine template spectrum. Although the original spectra cover a slightly wider wavelength range, we only process the region from 500~nm-1000~nm. The resolution of the spectra is $\lambda/\Delta\lambda \approx 10^6$ and the signal to noise in the continuum spans from 100-300 over the full wavelength range. For more information on the data and instrument setup, we refer the reader to \cite{lemke16} and \cite{Reiners16}. 

For generating the atlas, we choose a subset of spectra with a wide range in both solar radial velocity and airmass in order to achieve the best separation of each spectral component by leveraging the fact that the telluric component varies with airmass while the solar component varies with radial velocity. We opt to combine different days of data not only to maximize the range in these variables but also to leverage the different atmospheric conditions that will produce different telluric residuals and therefore reduce the possibility of confusing a telluric residual with a solar feature. We create 11 groupings of data that we perform our fits on with each group containing 12-15 spectra. Before running these fits we perform several pre-processing steps to the full sample of data that we describe below.

\begin{figure}
    \centering
    \includegraphics[width=0.99\linewidth]{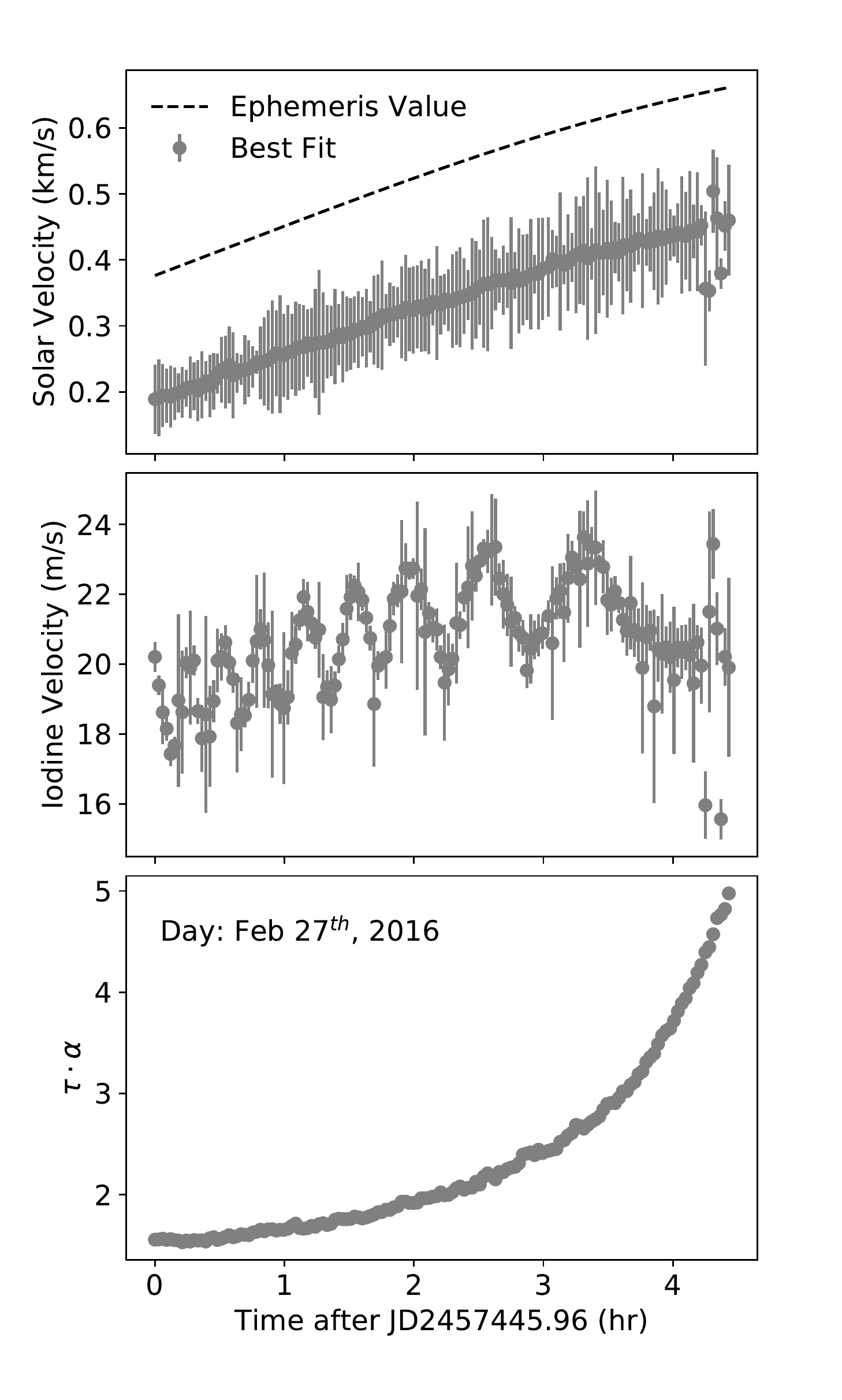}
    \caption{Observer-Sun velocity (top), iodine velocity (middle), and water vapor optical depth times airmass (bottom) versus time for observations taken on Feb. 27th, 2016. The dashed line in the top panel shows the actual solar velocity values that were calculated using the JPL web server and are offset from our measured velocities (gray points) due to the different zero-point velocities of the data and the template spectrum used to determine these values. The sequence of iodine velocities, measured by the iodine lines' positions relative to a template iodine spectrum, shows the drift in the Fourier Transform Spectrograph over this multi-hour observing run.}
    \label{fig:dnuvel}
\end{figure}

\paragraph{Flux Normalization}
We first prepare the data for fitting by dividing out the continuum to produce spectra with normalized flux levels. We do this individually for each spectrum using an automated process that steps through 150~cm$^{-1}$ subregions for each spectrum and records the maximum flux value and corresponding wavenumber value. A cubic spline is fit to these $\sim$70 points that describes the continuum. The raw flux data is divided by this spline description of the continuum in order to produce the final normalized flux data. Any errors in this process are accounted for in our ultimate fitting sequence by including a floating linear continuum correction. 

\paragraph{Pre-Fits: Solar and Iodine Velocities}
Using these flux-flattened spectra, we perform a fitting routine that determines (1) the solar velocity offset, v, relative to a template solar spectrum and (2) the wavelength calibration from the iodine lines using $\nu_c = \nu(1+\kappa)$, where $\nu_c$ is the corrected frequency array, $\nu$ is the original frequency array, and $\kappa=-\mathrm{v}_\mathrm{iod}/c$, where $\mathrm{v}_\mathrm{iod}$ is the iodine velocity, and $c$ is the speed of light. We determine these values a priori instead of optimizing them simultaneously with our telluric fitting sequence in order to enforce these values to be the same between the narrow (10~cm$^{-1}$) wavelength regions over which we perform each telluric fit. This ensures we are fully leveraging our knowledge of the locations of the stellar lines that is particularly important in regions that do not contain strong stellar features. The fitting routine we use to determine v and $\kappa$ for each spectrum follows the method described in Section 3 of \cite{lemke16} except instead of using an iodine-free IAG solar template, we use the telluric-corrected KPSA since we later use this for our starting guess of the solar spectrum in our full (solar+telluric) fits to the data. Although it is not necessary to begin with a solar model, this speeds up the iterative fitting process. 

In Figure \ref{fig:dnuvel} we show several parameters determined for a sequence of observations taken on Feb. 27th, 2016. In the top panel of Figure \ref{fig:dnuvel} we plot the measured velocities of the solar lines. The scatter in the measured solar velocity values is structured due to tracking errors, sunspots, and physical sources of line shifting that can occur (see \citealt{Reiners16} for more description). We ultimately shift our solar spectra by the Doppler velocity between the Sun and G\"{o}ttingen calculated using the JPL ephemeris generator\footnote{https://ssd.jpl.nasa.gov/horizons.cgi} (shown in Figure \ref{fig:dnuvel} as the black dashed line); therefore these effects do not affect the alignment of our final solar atlas and the random nature of the differences in the line shapes caused by tracking errors and sunspot position, for example, will be reduced due to averaging.

An example sequence of $\kappa$ values is shown in the middle panel of Figure \ref{fig:dnuvel}, where the error bars depict the standard deviation of the values measured for each of the ten 350~cm$^{-1}$ wide spectral regions that were fit independently then averaged to determine the final iodine velocity. The average uncertainty in the measurements for this day is 0.9~m~s$^{-1}$, which is typical of other days of observations. The oscillatory behaviour in $\kappa$ is also seen in the other observing runs and shows the intrinsic drift of the instrument. 

\paragraph{Pre-Fits: H$_2$O Optical Depth}
In Figure \ref{fig:dnuvel} in the bottom panel we show the product of airmass, $\alpha$, and water vapor optical depth, $\tau$, that accounts for the different column densities of water vapor between the various observations. The value of $\tau$ is first estimated from measuring the line depth of an isolated water vapor line located at 15411.73~cm$^{-1}$ and then further optimized in step 1 of our full fitting sequence that is described in \S \ref{sec:fit}. We only perform these optical depth fits for water vapor since oxygen is well-mixed in the atmosphere; therefore, using the airmass values of the observation to scale the oxygen telluric spectrum to the different observations is sufficient. This is described more in \S \ref{sec:telluric}.

\paragraph{Noise Determination} Quantization noise, photonic noise, and instrumental noise all contribute to the final noise in our FTS spectra. Of these, photonic noise dominates at high resolution where the informative component of the interferogram is small and can easily be swamped by the photonic noise due to a constant term related to the half power of the source \citep{ftsNoise10}. While in theory the noise of our measurements can be calculated, it is simple and accurate to deduce the final measurement noise by measuring the flux RMS of a portion of featureless spectrum. We do this for each observation by taking the noise for each spectrum to be the RMS of the flux normalized spectra over a 1.5~nm wavelength range starting at 1048.8~nm, which gives the noise in the continuum to be about 1\%. This region was chosen for its lack of solar and telluric features, but it must be noted that the actual noise levels vary slightly across the full wavelength span of the data due to the presence of telluric lines and variations in the sensitivity of the FTS. The RMS in the continuum drops to 0.3\% around 680~nm, and increases again at bluer wavelengths. Not accounting for the varying sensitivity of the FTS does not significantly affect the performance of our fits; however, it is important to modify our noise array over regions where the transmission drops due to saturated absorption lines. For this, we increase the noise determined for the continuum, $\sigma_\mathrm{cont}$, by a range of factors such that it is equal to 1.25$\sigma_\mathrm{cont}$ to 10$\sigma_\mathrm{cont}$ for where the transmission drops to 4.0-0.3\%, respectively. For example, in regions where the transmission is less than 1\%, we multiply the noise array by a factor of 2.5. The factors were chosen to match the observed noise in the data. We propagate the final noise array, $\sigma$, determined using the linear normalized flux, $\mathcal{F}$, to the noise of the logarithmic flux by $\sigma/\mathcal{F}$ and use these values in evaluating the optimization function for our fits.


\begin{table}[ht]
\centering
\caption{Summary of spectral parameters for each of the eleven fitting groups. For the observations included in each group we report the average value of the iodine velocity ($c \cdot \kappa$) in addition to the minimum and maximum values for the range of solar velocities, water vapor optical depths, and airmass values.}
\begin{tabular}{ccccc}
\centering
Group & $c \cdot \overline{\kappa}$ & v$_{min}$-- v$_{max}$ & $\tau_{min}$ -- $\tau_{max}$ & $\alpha_{min}$-- $\alpha_{max}$\\
No. & (m~s$^{-1})$ & (km~s$^{-1}$) &  & \\ \hline
0 & 4.1 & -0.1-- 0.3 & 1.0-- 5.3 & 1.1- 3.5  \\
1 & 6.0 & -0.5-- 0.4 & 0.3-- 1.3 & 1.1-- 3.0  \\ 
2 & 1.2 & -0.1-- 0.4 & 0.4-- 2.5 & 1.1-- 3.0  \\ 
3 & 9.4 & -0.7-- 0.4 & 0.2-- 1.0 & 1.1-- 3.1  \\
4 & 3.7 & -0.1-- 0.4 & 0.5-- 2.7 & 1.1-- 3.1  \\ 
5 & 8.8 & -0.7-- 0.4 & 0.3-- 1.7 & 1.1-- 3.2  \\ 
6 & 2.7 & -0.6-- 0.5 & 0.6-- 3.4 & 1.1-- 3.2  \\ 
7 & 9.4 & -0.5-- 0.4 & 0.4-- 2.2 & 1.3-- 3.4  \\ 
8 & 8.5 & -0.7-- 0.5 & 0.3-- 1.9 & 1.1-- 3.4  \\ 
9 & 6.9 & -0.6-- 0.3 & 0.4-- 1.6 & 1.2-- 3.4  \\
10 & 5.8 & -0.7-- 0.4 & 0.3-- 1.5 & 1.1-- 3.5  \\
\end{tabular}
\label{tab:params_summary}
\end{table}

\section{Modeling Methods}\label{sec:model}
Here we describe our model and justify our choices for how we represent each spectral component.

\subsection{Model Representation}
\begin{figure}
    \centering
    \includegraphics[width=0.99\linewidth]{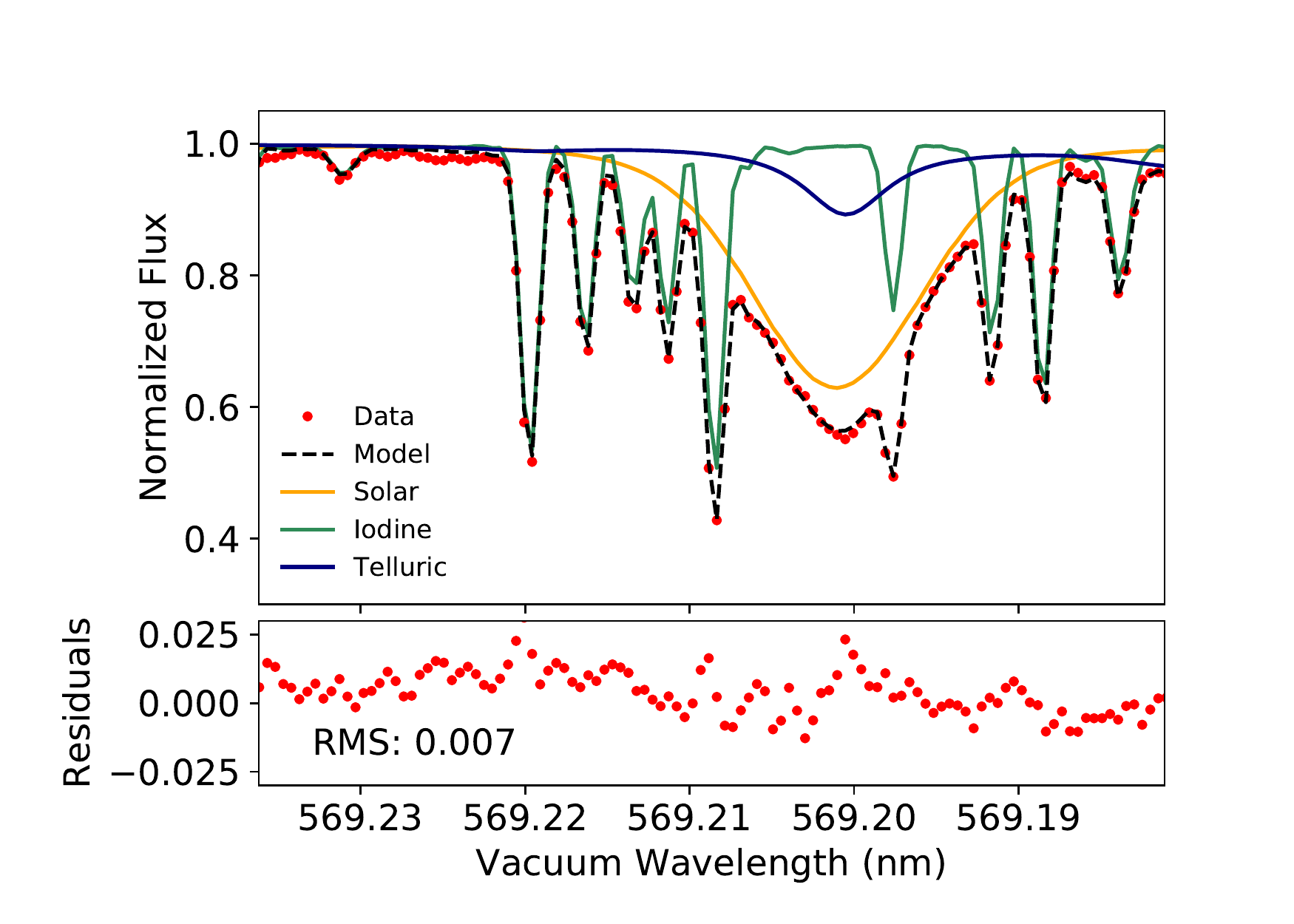}
    \caption{Example fit showing the various spectral components of our model. The continuum is not shown but is near unity for this region. The data are shown as red points with the model as the dashed black line that is equal to the product of the telluric (navy), solar (yellow), and iodine (green) spectra. The residuals (data minus model) are shown in the bottom panel.}
    \label{fig:components}
\end{figure}

To fit the IAG solar spectra, we construct a model that is composed of solar, telluric, and iodine spectral components in addition to a linear continuum model. We represent each component in units of absorbance for the fitting process. For computing reasons, we split each spectrum into 10~cm$^{-1}$ chunks that are fit separately. For each of the 11 groups of data, we simultaneously fit 10-15 spectra that range widely in the airmass of the observation and the Sun-observer velocity in order to best separate the telluric and solar components of the data. These groupings and their respective parameters are listed and described in Table \ref{tab:params_summary}. We therefore generate a model that is an N$_{\mathrm{spec}}$ by N$_\mathrm{points}$ array, where N$_\mathrm{spec}$ is the number of spectra being fitted and N$_\mathrm{points}$ is the number of data points in the fit region. Each model along the N$_\mathrm{spec}$ axis is generated using the same underlying solar and telluric spectral models, but is shifted and scaled according to the solar radial velocity and the species' column densities (including the airmass factor and $\tau$ for water vapor), respectively. Our final calculated model, $\mathcal{C}$, for each spectrum, indexed by $i$, can be represented as a sum of each component in units of absorbance:

\begin{equation}\label{eq:mod}
    \mathcal{C}(\nu)_i = \mathcal{A}_{\mathrm{T},i}(\nu) + \mathcal{A}_{\mathrm{S},i}(\nu) + \mathcal{A}_{\mathrm{I},i}(\nu) + \mathcal{A}_\mathrm{C}(\nu) \:,
\end{equation}

\noindent where we have used the subscripts T for telluric, S for solar, I for iodine, and C for the continuum and absorbance, $\mathcal{A}$, is just the logarithmic flux, $\mathcal{A} = -\log \mathcal{F}$. We demonstrate our model decomposed into all of these components in Figure \ref{fig:components} in which we have plotted the data, $\mathcal{O}$, as red points and our model, here $\mathcal{C}$, as the dashed black line, where both have been converted to linear flux units before plotting.

\subsubsection{Iodine Component}
For the iodine spectrum, we use the flux normalized template spectrum made from averaging together 450 iodine spectra taken with the FTS with a halogen lamp light source. The iodine template spectrum, $\mathcal{A}_\mathrm{I,temp}$, is shifted by the predetermined wavelength calibration, $\kappa$, that was found for each spectrum (see \S \ref{sec:data}) before being added to the final model using a cubic interpolation. There are therefore no optimization parameters related to the iodine model. We point out that we chose not to simultaneously solve for $\kappa$ in our full fits since the estimates from the wider wavelength regions in the pre-fits are more reliable than fitting for the iodine line shifts in each 10~cm$^{-1}$ chunk simultaneously with the other parameters. This is particularly true in regions where few or no iodine lines exist, which is around half of our spectral range. 

\begin{equation}
\mathcal{A}_{\mathrm{I},i} = \mathcal{A}_\mathrm{I,temp}(\nu \cdot (1+\kappa_i))
\end{equation}

\subsubsection{Continuum Model}
The continuum component of the model is important in order to capture errors in the flux normalization process that is a challenge in regions where the continuum is not well defined due to saturated telluric lines or a deep stellar feature. Because we are working in 10~cm$^{-1}$ chunks that are small compared to the curvature due to continuum modulation artifacts from the instrument, a linear trend sufficiently captures these errors. The continuum is specified in absorbance by two end point parameters, $\xi_l$ and $\xi_r$, and an extra vertical shift specified for each date, $\xi_\mathrm{date}$, so that 
\begin{equation}
    \mathcal{A}_\mathrm{C}(\nu)= \frac{\xi_l - \xi_r}{\nu_l - \nu_r} (\nu - \nu_l) + \xi_l + \xi_\mathrm{date}\: ,
\end{equation}
where $\nu_l$ and $\nu_r$ are the corresponding leftmost and rightmost wavenumber values. $\xi_\mathrm{date}$ is added to account for differences in the flux normalization process due to changing telluric absorption as well as differences in the illumination of the instrument that both result in slight continuum offsets between the spectra. These continuum differences are similar for observations taken on the same day so we only add an extra vertical shifts per unique observation date that is applied to the spectra taken on the respective day.

\subsubsection{Telluric Model}\label{sec:telluric}
In the spectral range of our data, O$_2$ and H$_2$O are the only species whose atmospheric abundances and absorption strengths result in signals that exceed the noise of our data. We therefore only include these two species and choose to model each individual line with a Lorentz profile. We use the High-resolution Transmission Molecular Absorption Database (HITRAN), version 2016 \citep{Gordon17} for a starting guess to the strength, $S_{0}$, width\footnote{We use $\gamma_{\mathrm{air}}$ from HITRAN for both molecules}, $\gamma$, and center position, $\nu_c$, and then optimize these parameters individually. We discuss our reasoning for pursuing this semi-empirical telluric model in Appendix \ref{sec:justify_tel}.

We optimize the Lorentz parameters for all individual line transitions with $S_0$ greater than $10^{-26}$ and $10^{-28}$ for water vapor and molecular oxygen, respectively, in HITRAN units of line intensity\footnote{https://hitran.org/docs/definitions-and-units/} (cm$^{-1}$/molecule/cm$^{-2}$). These parameters are queried directly from HITRAN using the HITRAN Application Programming Interface (HAPI; \citealt{hapi}). We also include in our model H$_2$O absorption features down to $S_{0}$=10$^{-28}$ cm$^{-1}$/molecule/cm$^{-2}$ that are not individually optimized, but do adopt a common modification on their line parameter values from an initial fitting step where all lines are shifted and scaled in unison. This line strength range corresponds to a linear flux approximately between 0.01-1\% deep with respect to the continuum for the highest airmass observations in this dataset. Different threshold values for each molecule are required due to their differing abundances in our atmosphere, $\Psi_\mathrm{mol}$, which are multiplied by each individually optimized $S$ value to determine the final line absorption strength. We summarize the full telluric model below, where we have indexed the H$_2$O lines by $l$ and the O$_2$ lines by $m$, where there are N$_{H_2O}$ water vapor lines and N$_{O_2}$ molecular oxygen lines in total. As before, each spectrum in the fit is indexed by $i$ and there are N$_\mathrm{spec}$ total. 
\begin{equation}\label{eq:telluric}
\begin{split}
    \mathcal{A}_{\mathrm{T},i}(\nu) = \tau_i \cdot \alpha_i \cdot \Psi_{\mathrm{H_2O}} \cdot \sum_l^\mathrm{N_{H_2O}} \mathcal{L}(\nu;\gamma_l,\nu_{c,l},S_l)  \\ + \alpha_i \cdot \Psi_{\mathrm{O_2}} \cdot \sum_m^\mathrm{N_{O_2}} \mathcal{L}(\nu;\gamma_m,\nu_{c,m},S_m) 
\end{split}
\end{equation}

\noindent For each species, we have one principle spectrum per day that is scaled to each solar observation by the airmass, $\alpha_i$, and for water vapor we additionally scale by the pre-fitted water vapor optical depth, $\tau_i$, previously determine for each observation. Although the telluric spectrum should be shifted by $k_i$, we do not include this since any bulk shift due to differences in instrument drift across spectra (\textless 10~m~s$^{-1}$ maximum, see column 2 of Table \ref{tab:params_summary}) will be small compared to a modification on $\nu_c$ (15 to 500~m s$^{-1}$) and will also ultimately be absorbed into the corrections on $\nu_c$. As will be discussed more in \S \ref{sec:fit}, the line parameters are first all modified simultaneously with all $\gamma$ values being multiplied by $f_{\gamma,\mathrm{mol}}$ for each molecule, the line strengths being multiplied by $\Psi_\mathrm{mol}$, and the line centers being shifted by $\delta_{air}\cdot P$, where $P$ is optimized and is physically motivated as a one dimensional pressure term and $\delta_{air}$ is the pressure-induced shift (units of cm$^{-1}$ atm$^{-1}$) and is provided in the HITRAN database for each line transition. 

\subsubsection{Solar Model}\label{sec:spline}
For our solar model, we use a cubic spline that we initialize to a flux normalized version of the telluric-corrected KPSA. The flux normalization for the KPSA is performed by simply dividing each 10~cm$^{-1}$ chunk by the maximum value in that spectral range. This performs well outside regions containing very wide stellar features; however, the continuum component of our model accounts for these offsets and is later used to correct for them. We describe this in Step 3 of \S \ref{sec:fit}.

Our spline is implemented using \texttt{BSpline} in the Python \texttt{scipy.interpolate} package and can be described as:

\begin{equation}\label{eq:spline}
    \mathcal{A}_\mathrm{S}(\nu) = \sum_{j=0}^{n-1} c_j B_{j,q;t}(\nu) \: .
\end{equation}

\noindent Here, we define the spline for a chunk of spectrum over which we define knot points, $t_j$. The final spline function can be written as the sum of coefficients, $c_j$, multiplied by each basis spline, $B_{j,q;t}$ that are defined in Appendix \ref{sec:bsplines}. 

For our application we use a cubic spline ($q$=3) and position knots at intervals of 0.1~cm$^{-1}$ in regions with low stellar absorption (\textless 10\% absorption) as determined by the Kitt Peak telluric-corrected solar atlas and use a knot spacing of 0.05~cm$^{-1}$ for stellar spectral regions with greater than 10\% absorption. We found that this knot sampling was able to capture the curvature of the spectral features and a coarser spacing of knot points would introduce oscillatory numerical features above the noise in regions with high curvature. Because each knot point has multiplicity one (no overlapping points), our resultant stellar spectrum will be smooth, as desired. In our fits, we optimize the coefficients, $c_j$, of the spline that are initialized by performing a least squares minimization between the spline and the telluric-corrected KPSA.

The final stellar model array, $\mathcal{A}_{\mathrm{S},i}$, contains N$_\mathrm{spec}$ stellar models each shifted by the solar and iodine velocity already measured for each spectrum included in the fit. We generate $\mathcal{A}_{\mathrm{S},i}$ by simply evaluating our solar spline at the array of wavenumbers modified by $\kappa_i$ and v$_i$:

\begin{equation}\label{eq:solar}
    \mathcal{A}_{\mathrm{S},i}(\nu) = \mathcal{A}_\mathrm{S}(\nu\cdot(1+\kappa_i)(1-\mathrm{v}_i/\mathrm{c}))
\end{equation}

\begin{table}\label{tab:parameters}
\centering
\caption{A summary of the optimization parameters for our model described in the text.}
\begin{tabular}{ccc}
 Model  & Optimized  & Prefit    \\ 
 Component &  Parameters &  Parameters   \\ \hline
 $\mathcal{A}_\mathrm{S}$    &  $c_j$             &   v$_i, \kappa_i$       \\
 $\mathcal{A}_\mathrm{T}$    & See Table \ref{tab:tel_pars} &   $\tau_i$       \\
 $\mathcal{A}_\mathrm{C}$    &  $\xi_l$, $\xi_r$, $\xi_{\mathrm{date}}$ &  -  \\
 $\mathcal{A}_\mathrm{I}$    &      -         &   $\kappa_i$     \\
\end{tabular}
\end{table}

\begin{table}
\centering
\caption{List of telluric model parameters relevant to select absorption features depending on their transition line strengths. We additionally denote the line strength cutoff for weak telluric features that are omitted from the model and additionally define the minimum line strength defining the boundary of the `strong' group of features that are fit together with the solar spline model in Step 2 of the fitting sequence (see text for more description).}
\begin{tabular}{ccc}
 Species  & Line Strengths  & Optimized Parameters    \\ \hline
H$_2$O    &  S \textless 10$^{-28}$           &  (omitted)      \\
          &  S \textgreater 10$^{-28}$  &  $f_{\gamma,H_2O}$, $\Psi_{H_2O}$, $P$       \\
          &  S \textgreater 10$^{-26}$   &  $\gamma_l$, $S_l$, $\nu_{c,l}$   \\
          &  S \textgreater 10$^{-25}$   &   (`strong' lines) \\
          & &  \\
O$_2$     &  S \textless 10$^{-28}$           &  (omitted)      \\
          &  S \textgreater 10$^{-28}$  &  $f_{\gamma,O_2}$, $\Psi_{O_2}$, $P$, $\gamma_m$, $S_m$, $\nu_{c,m}$      \\
          &  S \textgreater 10$^{-27}$   &  (`strong' lines) \\
\end{tabular}
\label{tab:tel_pars}
\end{table}

\section{Fitting \& Processing Steps} \label{sec:fit}
Here we discuss the fitting sequence in detail and describing how we generate the final solar atlas in addition to the telluric spectra extracted from the data set.

\subsection{General Fitting Routine}

\paragraph{Step 1: Telluric Optimization}
The optimization sequence begins by stepping through the spectra in 10~cm$^{-1}$ chunks (each contains 664 datapoints) and for each chunk fitting $N_\mathrm{spec}$ spectra with the solar spectral model set to the KPSA for that region while we iteratively optimize\footnote{We found best behaviour using the sequential least squares programming (SLSQP) algorithm in the \texttt{scipy.optimize.minimization} function} the continuum and telluric parameters. For this we minimize a $\chi^2$ term with a penalty term, $p_1$, added in order to encourage the model to go to zero in saturated regions that otherwise do not contribute significantly to the $\chi^2$ value due to the large flux uncertainties. If $\chi_1^2$ is the objective function for Step 1, it can be summarized as $\chi_1^2 = \chi^2 + p_1$ with $\chi^2$ and $p_1$ defined as follows.

\begin{equation}\label{eq:chi2}
    \chi^2 = \frac{1}{N_\mathrm{spec} N_\mathrm{point}}\sum_\nu \sum_i \frac{[\mathcal{O}_i - \mathcal{C}_i]^2}{\sigma_i^2/\mathcal{F}^2_i},
\end{equation}

\begin{equation}
    p_1 = \frac{1}{N_\mathrm{sat}}\sum_{i}\sum_{\nu_{\mathrm{sat}}} \frac{[e^{-\mathcal{O}_i} - e^{-\mathcal{C}_i}]^2}{\sigma_\mathrm{med}^2} 
\end{equation}

\noindent Here we have denoted the median uncertainty over each spectrum as $\sigma_\mathrm{med}$ that provides a scaling according to the noise in each spectrum without diminishing the term as would happen if we used the $\sigma$ values, which are large over those saturation regions.

For the iteration series, we begin by optimizing the continuum parameters, $f_\gamma$ and $\Psi$ for each species, and $P$ that scales the linecenter shifts. Lines with strength greater than 10$^{-28}$ for both species are included in the model and are modified based on the best fit values of $f_\gamma$, $\Psi$, and $P$, which effectively modifies all features in unison. This unified shift and scaling serves as a first approach to the best fit solution and is additionally useful for the weakest features that have line depths near the noise making them challenging to fit individually. In the next stage, we optimize the continuum and telluric models, where all three parameters for each line of the telluric model is allowed to vary individually. As recorded in Table \ref{tab:tel_pars}, we optimize the line parameters for individual lines with strengths greater than 10$^{-26}$ and 10$^{-28}$ for water vapor and oxygen, respectively. We split the water vapor lines into two groups based on line strength that may be fit separately to improve computation times: these are a `weak' group and a `strong' line group. The lines in the `strong' group\footnote{The strong lines have depths greater than around 5\% in linear normalized flux on average.} includes absorption features with $S$ ten times the lower thresholds just defined for individually fitted H$_2$O lines with the `weak' group containing the remaining features ($10^{-26} < S < 10^{-27}$). We iterate between fitting these two telluric groups and the continuum parameters until convergence (i.e. no significant changes in the $\chi_1^2$ value). At the end we allow both groups of telluric lines and the continuum to vary simultaneously, having kept $f_\gamma$, $\Psi$, and $P$ fixed in this iteration process. We note that we bound the amount that the lines centers can shift to 0.01~cm$^{-1}$ in each optimization stage to avoid the fits swapping features in the data.

We find that our fits perform well but that the optical depths of the water vapor lines are poorly estimated by our initial effort that determined $\tau$ by fitting one isolated water vapor line. We improve this by iterating between fitting for $\tau_i$ over all spectra and performing the fitting sequence described here until convergence. We only do this for one 10~cm$^{-1}$ wide range of spectra starting at 11010.0~cm$^{-1}$ (907.44 - 908.26~nm) for which the region is dominated by deep water vapor lines. We find the wings of the lines are very sensitive to the value of $\tau$ and fitting multiple lines at once significantly improves our estimates of the optical depth for each spectrum.

\paragraph{Step 2: Solar and Telluric Fitting}
Using the best fit estimates for the continuum and telluric line parameters, we now optimize both the spline coefficients and the group of strong telluric transitions. We choose not to fit the weak group simultaneously with the spline optimization since these lines are fit reliably well in Step 1. To perform this fit we use the optimization function from Step 1 and add another penalty term that serves to penalize the fit for adding features to the solar spectrum over saturated regions. This is to alleviate the potential problem of the spline filling in saturated regions, where no photon information exists. The two penalty terms together therefore ensure that the core of a saturated line is fit properly (i.e. the model goes to zero) while avoiding the scenario where a narrow, deep feature in the spline fills in the core, since this is unlikely and in any case cannot be constrained. We choose not to completely prevent solar features in these regions since a solar line will often overlap the edges of saturated regions and we wish to not compromise the shape of these features by abruptly forcing the spline, that must be smooth, to zero. We therefore define the penalty, $p_2$, as

\begin{equation}
    p_2 = \beta  \sum_{\nu_{sat}} \mathcal{A}_{\mathrm{S}} \: ,
\end{equation}

\noindent where the coefficient $\beta$ is a scaling factor that adjusts the penalty term to an effective range that does not force the solar spectrum to zero, but still prevents large, unnecessary additions to the spectrum. The objective function for step 2 can be summarized as ${\chi_2}^2 = \chi^2 + p_1 + p_2$. Because we have a good first guess to the solar spectrum and our $\chi^2$ value, tuning $\beta$ uniquely to each portion of spectrum can be done based on this prior information. Since our $\chi^2$ value is in theory at best unity when the residuals only encompass noise on the magnitude of the noise of the data, we can define $\beta$ to be the value such that the initial penalty term is 0.5 or around 50\% of $\chi^2$ for a well optimized fit. We therefore set $\beta = \frac{0.5}{\sum_{\nu_{sat}} \mathcal{A}_{\mathrm{S},0}}$, where $\mathcal{A}_{\mathrm{S},0}$ is the initial solar spectrum that has been set to the KPSA. This method of tuning a penalty term is typical in cases where overfitting can potentially be an issue (e.g. \cite{intro_bed19}).

\begin{figure}
    \centering
    \includegraphics[width=0.99\linewidth]{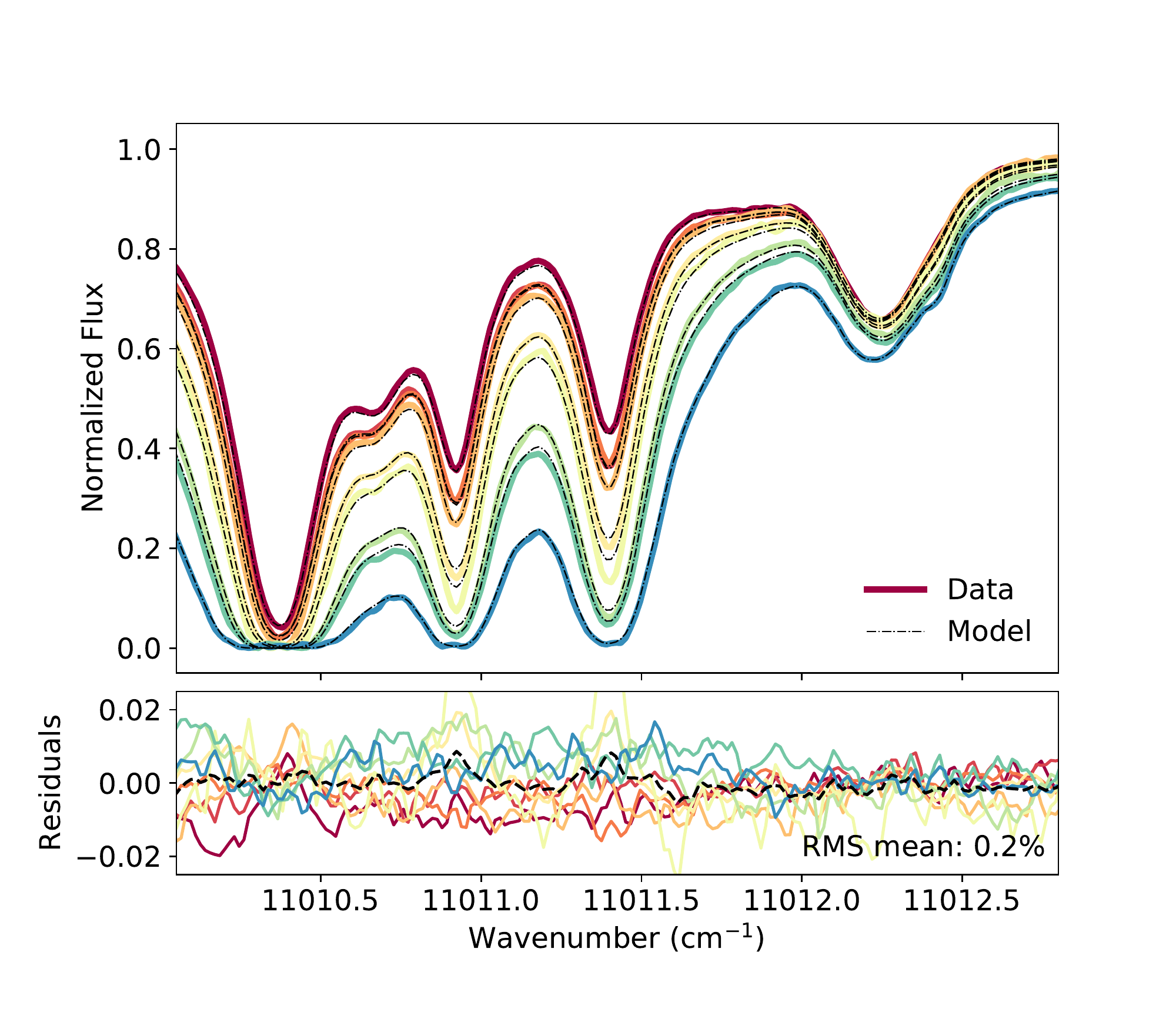}
    \caption{Example fit shown for a small wavelength region for one group of spectra. The flux-normalized spectra are shown with colors corresponding to the water vapor optical depth and the best fit models for each are plotted as a dashed black line (top). The residuals for each fit is shown (bottom) with the same colors. A solar feature on the right is apparent due to its lack of change with airmass and demonstrates the small Doppler velocities that cause line shifts smaller than a fraction of a line width.}
    \label{fig:eg_fit}
\end{figure}

\paragraph{Step 3: Correcting the Continuum}
As noted before, the fitting process is performed for each 10~cm$^{-1}$ subregion separately. Before removing the continuum, telluric, and iodine solutions to extract the final solar spectrum, we must address the fact that the continuum solution contains the corrections to both the data and solar spectral continua. We find that in saturated telluric regions, the original continuum corrections are due to errors in the flux normalization done to the data, while the solar spectrum normalization process only fails for regions where the entire spectral chunk contains a wide solar feature that spans the 10~cm$^{-1}$ subregion since otherwise the solar model is at maximum (unity) between stellar lines. Over our wavelength range, there are five occurrences of a wide solar feature and these all are present in regions that do not overlap saturated telluric features. We utilize this in separating the continuum offsets originating from the stellar model from offsets due to the continuum correction performed to the data.

To remove the offsets in the continuum solution due to errors in the initial normalization to the solar spectrum, we take the final best-fit continuum array and, if the spectral subregion under consideration is overlapping a dense telluric band, we leave the continuum alone. If the subregion does not contain dense\footnote{We use an arbitrary definition for what constitutes `dense' that involves summing the telluric model and comparing it to a predetermined threshhold value.} telluric features, we modify the best-fit continuum model by subtracting from it the median value of the continuum solutions determined for each spectrum in the group. Because we will later use the solar spline model to replace regions containing telluric residuals, we add these subtracted values to our stellar spline model. This transfers any continuum offsets originating from the stellar normalization process back to the stellar spectrum, while keeping information about the relative offsets between the spectra grouped by date. Typically these offsets are small over telluric-free regions.

\subsection{Generating the Final Solar Spectrum}
\paragraph{Removing Best-Fit Model Components}
Using the corrected continuum array along with the telluric model and iodine template, we can subtract these from the FTS data to leave just the solar component:

\begin{equation}
    \mathcal{F}_{\mathrm{S},i} = e^{ - (\mathcal{A}_{\mathrm{data},i} - \mathcal{A}_{\mathrm{I},i} - \mathcal{A}_{\mathrm{T},i} - \mathcal{A'}_{\mathrm{C},i})} .
\end{equation}

\noindent Here we have denoted $\mathcal{A'}_{\mathrm{C},i}$ as our corrected continuum array and $\mathcal{F}_{\mathrm{S},i}$ as our final solar spectra for each observation defined over our full wavelength span and we have converted from absorbance to transmission. We note that in regions where the data are saturated there will be spurious values due to dividing regions dominated by noise by the model that is approximately zero in those regions. Residuals from the telluric subtraction may also remain in the final solar spectrum and are typically visible over deeper lines (\textgreater 10\% absorption). Instead of excising these regions entirely we instead replace them by our spline model. These regions are flagged in the final spectrum. 

\paragraph{Velocity Zero Point Determination}
To achieve an absolute wavelength calibration, we use the iodine catalog from \cite{iodine_ascii} who recorded a Doppler-limited iodine spectrum using an FTS and corrected the wavelength scale to match other wavelength-calibrated iodine atlases including that of \cite{GERSTENKORN1981322}. From their comparison to these other atlases, they estimate that their spectrum is reliable to $\pm$0.003~cm$^{-1}$ across their frequency range of 14250-20000~cm$^{-1}$. We use their recorded spectrum to find the offset of our template iodine spectrum and find that our iodine spectrum is shifted redward of the template by about 70~m~s$^{-1}$, or 0.004~cm$^{-1}$. We shift our final solar spectrum by 70~m~s$^{-1}$, such that $\kappa$ = 3.2$\times 10^{-7}$ and adopt the uncertainty in the template iodine spectrum of $\pm$0.003~cm$^{-1}$, which translates to $\pm$45~m~s$^{-1}$ at frequencies of 10000~cm$^{-1}$ and $\pm$90~m~s$^{-1}$ at 20000~cm$^{-1}$. We note that any shifts in our iodine spectrum due to a temperature difference between our setup and the temperatures used by \cite{iodine_ascii} should be over 10 times smaller than our adopted uncertainties for the absolute shift of our final stellar spectrum \citep{iodine_stability}. We additionally attempted to derive a zero point offset solution using the central positions of the telluric lines as compared to their catalog linecenter modified by the pressure-induced line shift. This produced a consistent result but was less precise. This is described more in Appendix \ref{sec:zpo}.

\begin{figure*}
    \centering
    \includegraphics[width=0.98\linewidth]{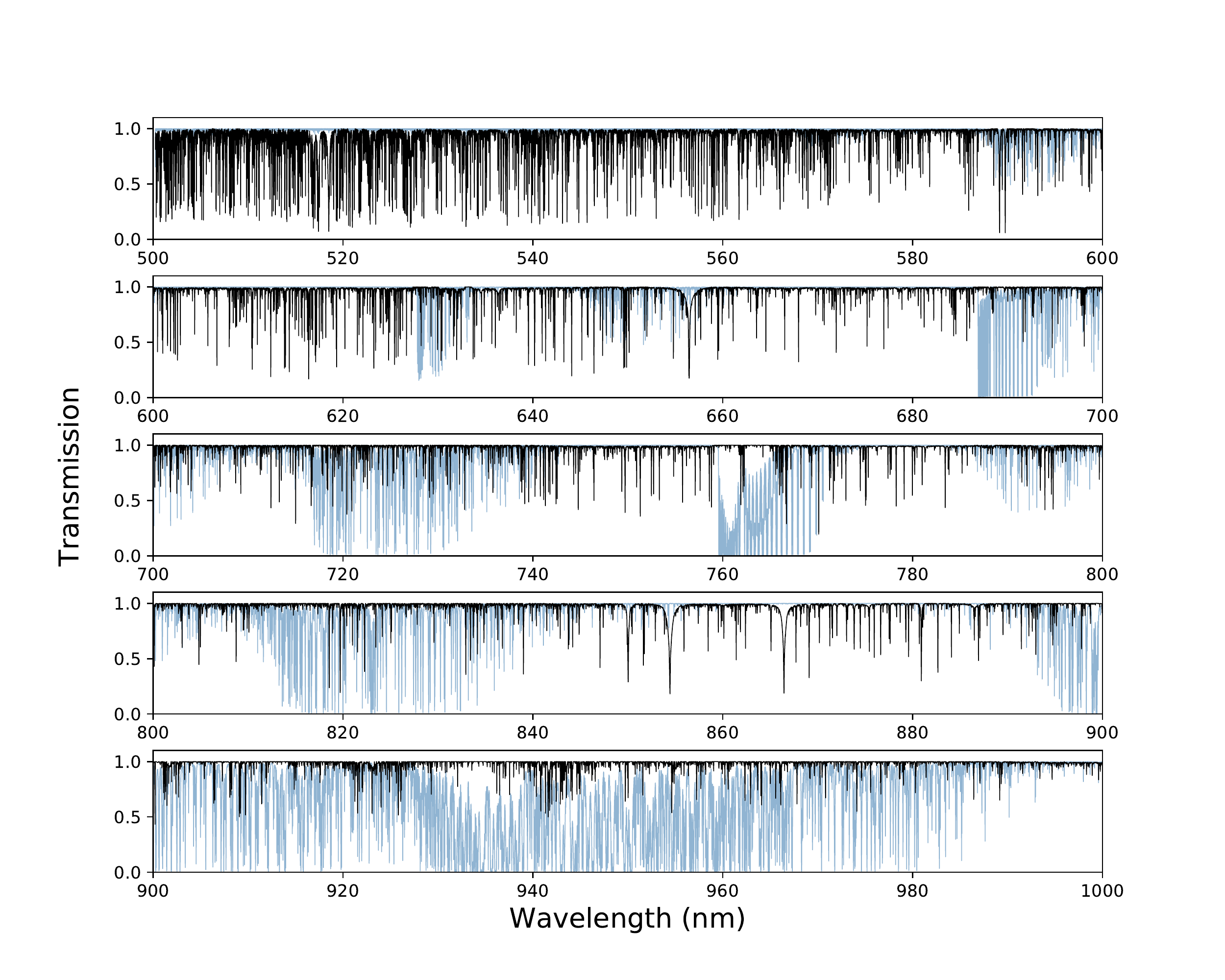}
    \caption{Transmission as a function of wavelength for the full telluric-corrected IAG solar atlas. In black is the final solar spectrum and in blue is an extracted telluric spectrum. The telluric model shown is typical of conditions at G{\"o}ttingen (precipitable water vapor of $\sim$10~mm)}. 
    \label{fig:full_spec}
\end{figure*}

\paragraph{Combining Spectra}
We combine our spectra after shifting $\mathcal{F}_{\mathrm{S},i}$ according to the velocity v$_{\mathrm{eph},i}$ between G{\"o}ttingen and the Sun at the time of the observation. We also perform a second shift for the calibration velocity measured from the iodine lines and the final shift for the absolute zero point velocity. We combine the spectra by stepping through the same 10~cm$^{-1}$ chunks and, before averaging, remove spectra with extreme residuals due to either a poor fit or a large airmass value that, although are useful for constraining the spline fit, have higher uncertainties and the least information over telluric regions. We record the final average of the remaining spectra as our telluric-free IAG solar flux atlas and record the standard deviation of these remaining spectra as the final uncertainty. We plot the final atlas in Figure \ref{fig:full_spec}. 

For ease of use and since the uncertainty array will fail over telluric-contaminated regions replaced by the spline model, we create a flag array for the final solar spectrum to identify regions with varying levels of telluric absorption: 0 indicates a robust spectral region, 1 indicates a region with telluric absorption exceeding 10\%, 2 for telluric absorption exceeding 25\%, and 3 for saturated regions. Of these, flags greater than or equal to 1 correspond to regions that have been replaced by the spline model.

We inspect the final solar atlas and notice that the oxygen A bandhead and the area around the HeNe laser used for internal wavelength calibration of the FTS both contain spurious features. We excise these two regions (14522.0-14523.6~cm$^{-1}$ for the O$_2$ residuals and 15795.1-15799.0~cm$^{-1}$ for the HeNe residuals) by replacing them with unity and assigning zeros to the uncertainty array and a flag value of 3 for the full extent of both regions. We also point out that several deep solar lines in the 500-555~nm (18000-20000~cm$^{-1}$) range were very close to saturated such that, if an iodine feature overlapped the deepest portion, the final division of the iodine spectrum would be leave a large residual in the solar spectrum. Since the iodine features are stable in time, the solar spline model occasionally fit the residuals such that the model could not be reliably used to replace the erroneous spectral shape. A few near-saturated solar lines therefore contain poorly constrained line core shapes, however, the final uncertainties capture the magnitude of these deviations.

\subsection{Extracting Telluric Spectra}
The high resolution and high signal to noise of the IAG solar spectra makes it a good data set for various telluric line studies. For example, this may include studying commonly used telluric modeling codes, the stability of oxygen lines, and the impact of micro-telluric lines on radial velocity measurements. We therefore create solar-corrected telluric spectra from the data set and also make these publicly available for future studies. To do this, we divide the linear flux normalized data by the shifted iodine spectrum and by the final stellar model shifted by the solar velocity determined from the pre-fits. We then shift each spectrum by its iodine velocity, $\kappa$, and the zero point velocity and then save each spectrum recorded with the airmass and our measured water vapor optical depth values for the observation. We recommend the solar spectrum be downloaded and referred to as well depending on the use of the telluric spectra since overlapping solar lines could potentially skew the shape of an extracted telluric spectrum.

\hfill \break
\noindent We make available the solar and telluric data products online\footnote{http://web.sas.upenn.edu/ashbaker/solar-atlas/  or Zenodo, DOI:10.5281/zenodo.3598136}. In Figure \ref{fig:full_spec} we show the final solar atlas covering 500-1000~nm in black with an example telluric spectrum extracted from the data in blue.

\begin{figure*}
    \centering
    \includegraphics[width=0.7\linewidth]{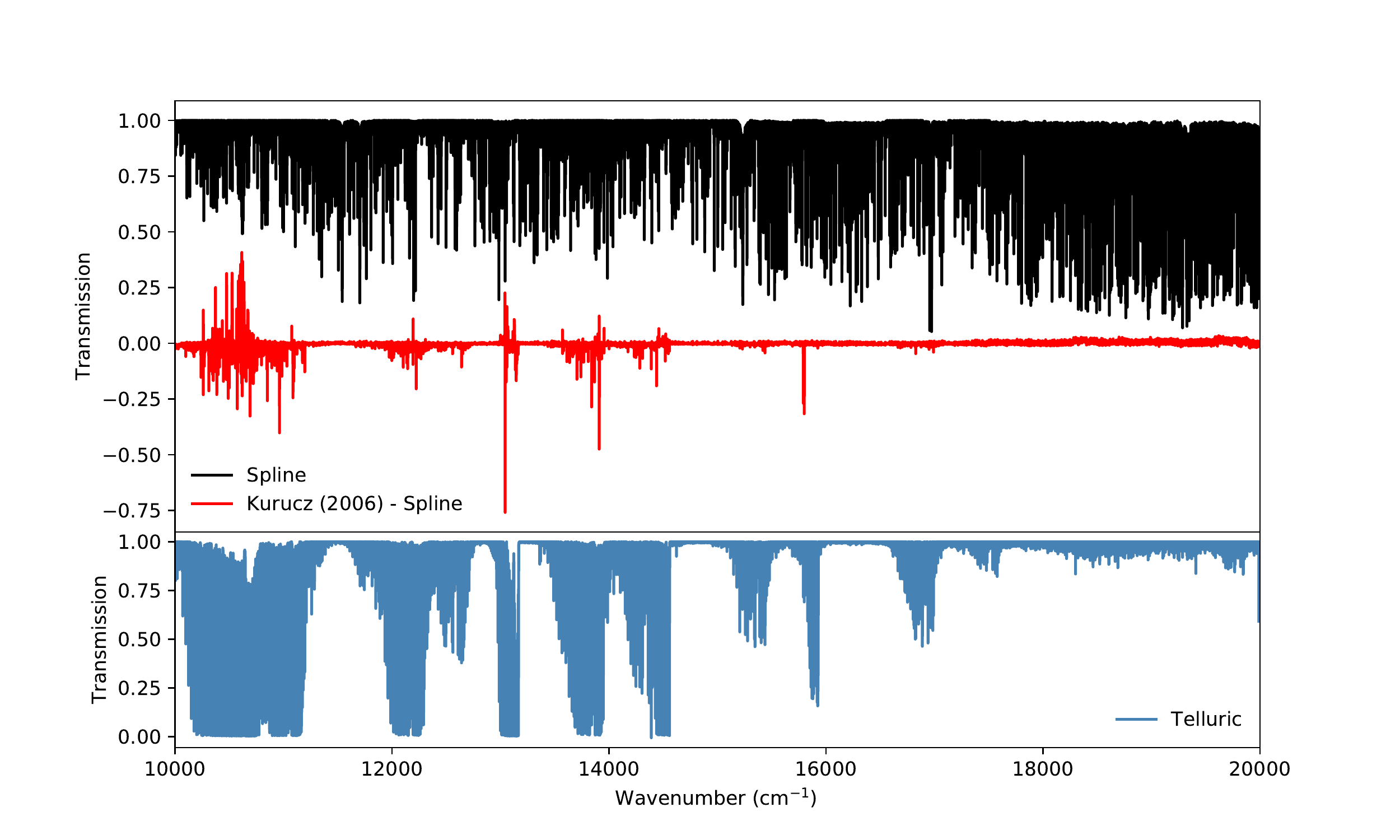}
    \caption{Comparison of Kitt Peak and IAG telluric-corrected solar atlases. The top panel contains the IAG best fit solar spline model (black) and the residuals between \cite{Kurucz06} and the spline (red). In the bottom panel we show a telluric model in blue.} 
    \label{fig:compare_KPSAfull}
\end{figure*}

\begin{figure*}
    \centering
    \includegraphics[width=0.9\linewidth]{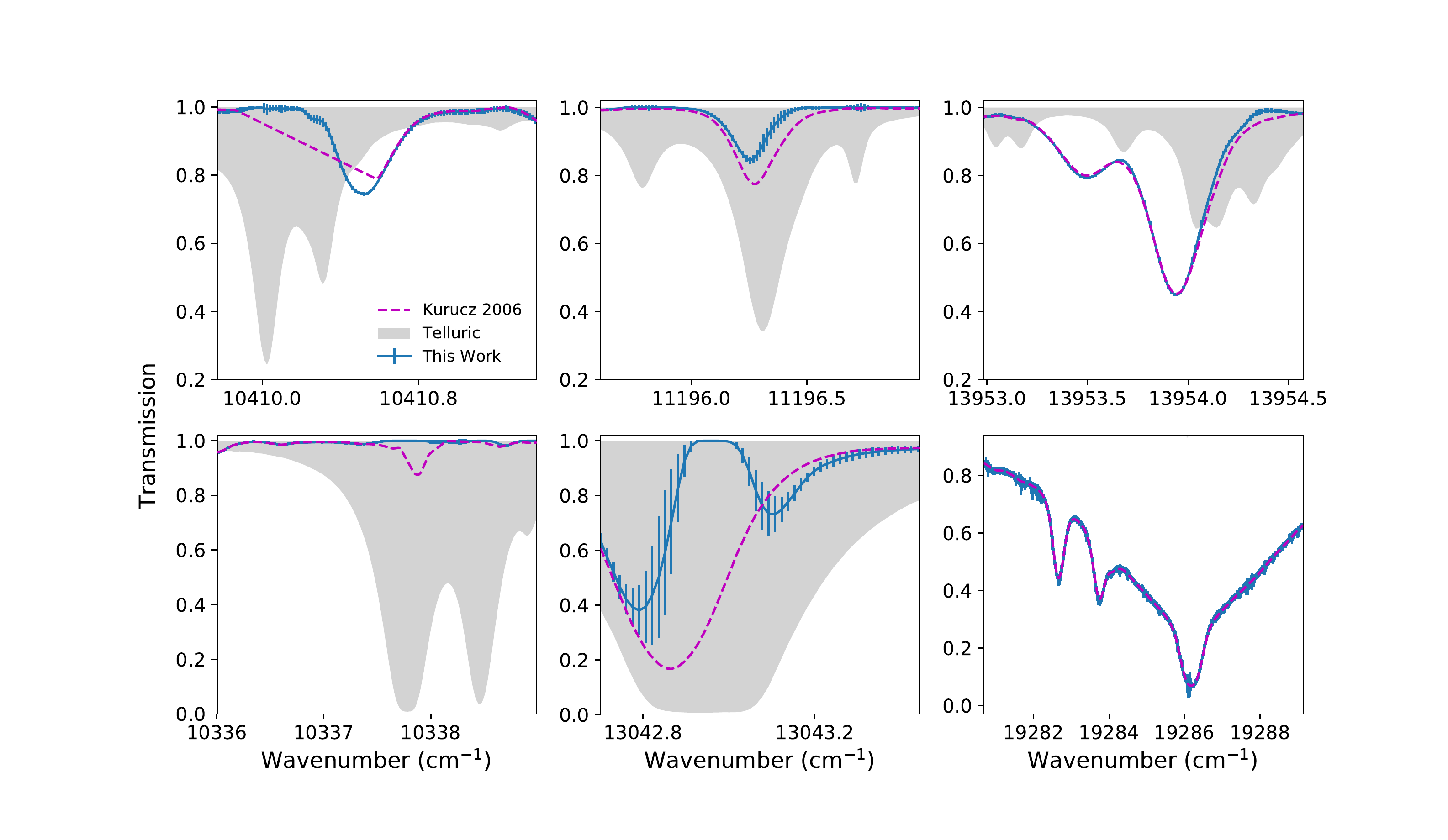}
    \caption{Comparison of Kitt Peak and IAG telluric-corrected solar atlases over select wavelength ranges. Uncertainties are generated by taking the standard deviation of all telluric-subtracted spectra used to generate the IAG solar flux atlas.}
    \label{fig:compare_KPSA}
\end{figure*}

\section{Analysis \& Discussion}\label{sec:analyze}
Here we compare our final solar atlas to the KPSA and discuss our telluric fits. We additionally compare our best fit telluric model parameters to the starting HITRAN values and comment on several observations.

\subsection{Comparison to the Kitt Peak Atlas}
We compare our final spectrum to the KPSA that we used as a starting guess to our solar spline model and note that the bulk of the differences occur over dense telluric bands, as expected. This can be seen in Figure \ref{fig:compare_KPSAfull} in which we plot the residuals between the two solar spectra. In regions overlapping telluric features \textless50\% in depth, we observe that the differences are \textless3\%. Spectral features overlapping saturated or near-saturated lines occasionally differ as high as 25\% or more. Inspecting the differences between the two spectra over strong telluric lines shows (a) similarly identified features differing in line strength and/or shape and (b) solar features present in one atlas but not in the other. We show examples of these cases in Figure \ref{fig:compare_KPSA}.

Most of the differences in the line shape when the solar line is bordering a saturated feature, over which we have no information and our algorithm encourages the line to be narrower, sometimes splitting the line in two (e.g. bottom middle of Figure \ref{fig:compare_KPSA}). This was expected and the saturated flag we provide can be used to ignore the erroneous section of the line. We do see that sometimes our solution finds a narrower line over non-saturated lines (top middle and right of Figure \ref{fig:compare_KPSA}) than in the KPSA, though we note the reverse case happens as well. 

In Figure \ref{fig:compare_KPSAfull}, where we plot the residuals between the two spectra, we notice that more residuals fall below zero corresponding to the KPSA typically having lower transmission than our spectrum in discrepant regions. A partial explanation is that our algorithm will return to the continuum in saturated regions where there is no information otherwise. Also, occasionally \citealt{Kurucz06} would replace regions that had remaining large residual features with a linear interpolation that connects the adjacent regions. Our spline in contrast would smoothly return to the continuum level (e.g. top left of Figure \ref{fig:compare_KPSA}).

In the blue region of our spectrum (wavenumbers higher than 17500~cm$^{-1}$) it can be seen that there is more scatter in our solution compared to the KPSA. This is due to higher instrument noise in this region as well as iodine features that were poorly removed. Because the iodine cell was not temperature stabilized at the time these data were observed, the line strengths of the iodine lines in the template spectrum differed slightly from the iodine strengths in the data. Nevertheless, the uncertainties in the final solution account for this (e.g. bottom right of Figure \ref{fig:compare_KPSA}).

\subsection{Missing Water Vapor Lines}
Three prominent spectral features were found that were unaccounted for in our model. For each feature, we correlated the integrated line absorption with airmass and determined that all are telluric in origin. Furthermore, we find that each has a one-to-one correlation to the integrated strength of a water vapor feature of similar depth, which confirms that all are water vapor lines. We add these lines to our local HITRAN water vapor database before performing the fitting sequence in the these regions. We initialize the fitting parameters for each line to those of lines similar in strength and summarize these values in Table \ref{tab:missing_lines}. We queried the HITRAN 2016 database around the line centers but did not find any other possible candidate species having large enough strength to explain the features. We note that the HITRAN line lists are extensively complete, especially for water vapor over optical wavelengths, and so it is possible these lines were accidentally omitted between versions as we see no reason that these lines would be missed in the detailed laboratory experiments that source the HITRAN line lists. These three lines were the only ones found missing, although we estimate that we would be limited in our ability to detect missing lines weaker than about 0.5\% in the continuum and about 2\% over other features, since this is on the order of the residuals in the continuum and over some telluric lines, respectively. Additionally, we sometime find that some dense saturated regions are easily fit very well, while some regions leave larger, structured residuals that could be due to a missing telluric feature or an overlapping solar line, but we do not have the ability to determine the true cause.

\begin{table}[ht]
\centering
\caption{Telluric lines found unaccounted for from our HITRAN 2016 input parameters.}
\begin{tabular}{cccc}
 Line Center  & Initialized Strength & Initialized Line Width  \\ 
 cm$^{-1}$ & cm$^{−1}$/(molecule cm$^{−2}$) & cm$^{-1}$/atm \\ \hline
  10519.8  & 4$\cdot 10^{-24}$     &     6.3$\cdot 10^{-2}$    \\
  13941.54 & 1.76$\cdot 10^{-24}$  &     8.8$\cdot 10^{-2}$   \\
  13943.0  & 1.76$\cdot 10^{-24}$  &     8.8$\cdot 10^{-2}$   
\end{tabular}
\label{tab:missing_lines}
\end{table}

\begin{figure*}
\centering
\begin{minipage}{.47\textwidth}
  \centering
  \includegraphics[width=0.99\linewidth]{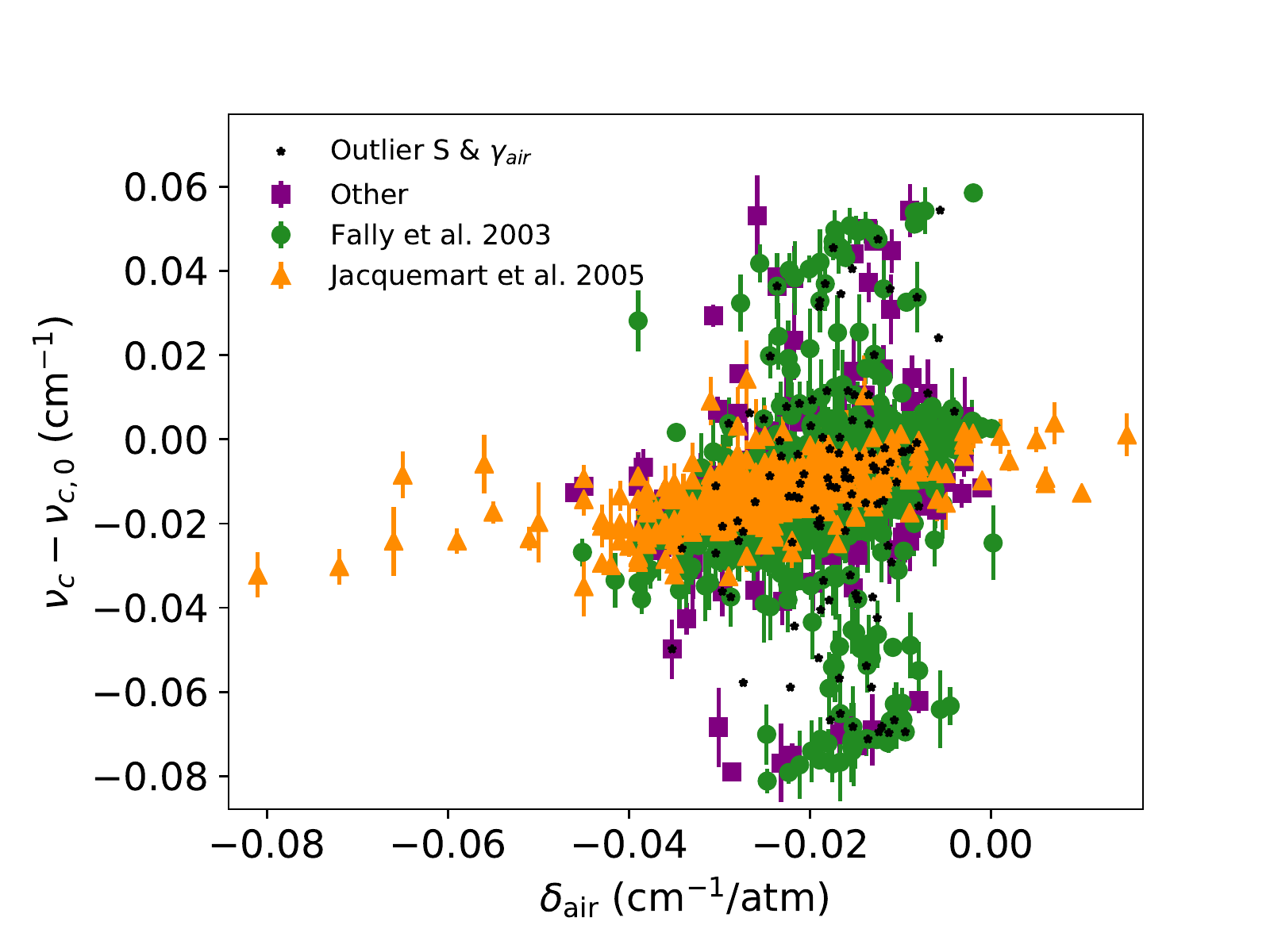}
    \caption{Best fit line centers minus the HITRAN value for H$_2$O as a function of $\delta_{air}$, the pressure induced line shift. Data points are colored by select references. The data are averaged together from separate fits and points with errors higher than 0.01~cm$^{-1}$ are removed for clarity. Black stars indicate independently selected points that have outlier line strengths and widths in comparison to the HITRAN values.}
    \label{fig:tel_shift}
\end{minipage}
\hfill
\begin{minipage}{.47\textwidth}
  \centering
  \includegraphics[width=0.99\linewidth]{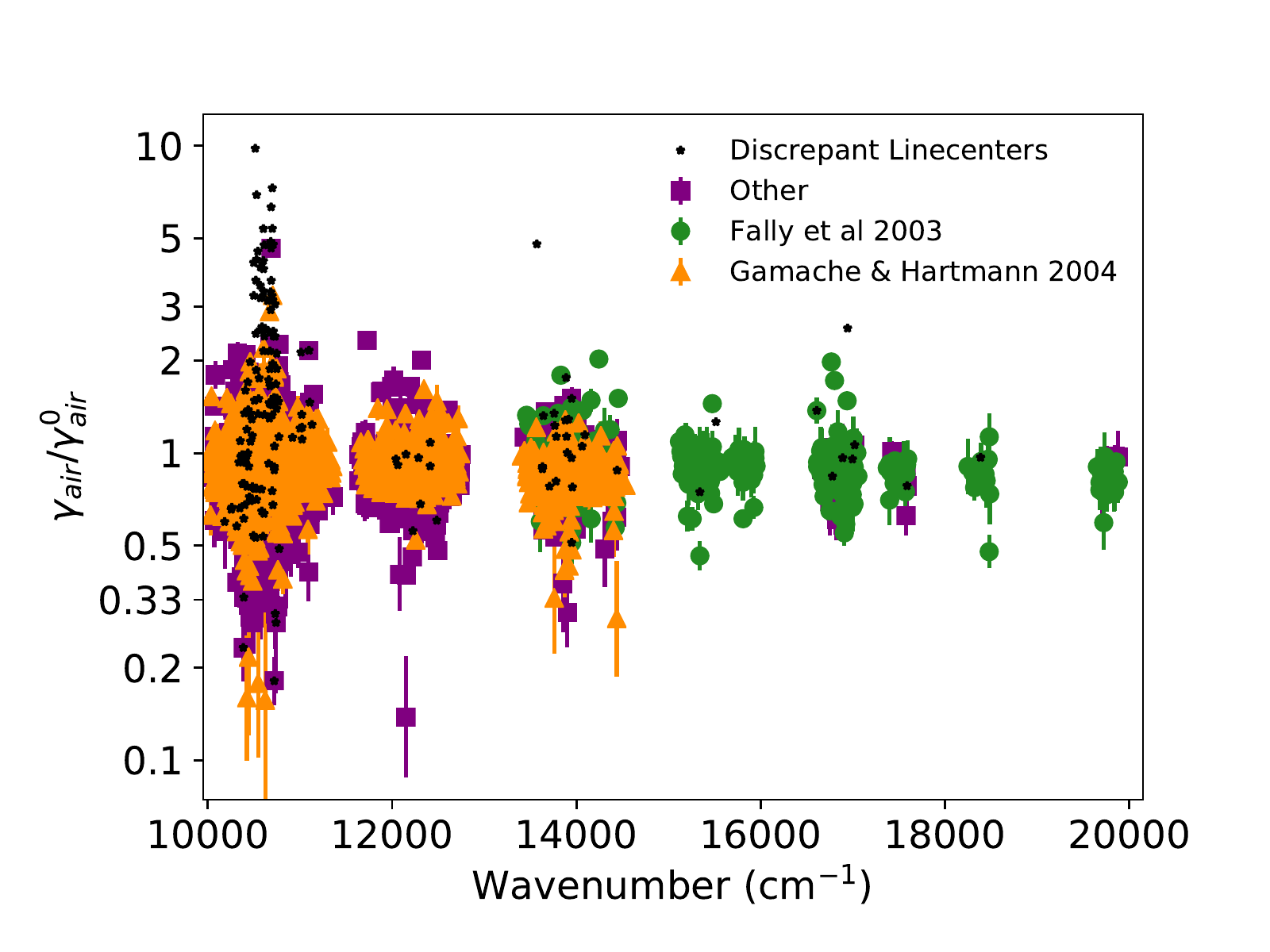}
    \caption{Best fit Lorentz widths over the original HITRAN database air broadened $\gamma$ value plotted versus wavenumber for H$_2$O and colored by select references. Data points are the average from several fits and points with statistical uncertainties greater than 2\% are removed. The black stars are independently selected lines that have discrepant best-fit linecenters.}
    \label{fig:telbad_nu}
  \end{minipage}
\end{figure*}

\subsection{Comparison to HITRAN}
The HITRAN line lists are a vital resource to many scientific studies from modeling Earth's atmosphere to remote detection of a molecular species. The water vapor line lists are of particular importance due to the role water plays in Earth's atmosphere and its large absorption features across the optical to NIR spectrum. A large amount of theoretical and laboratory work has gone into improving these line parameters, particularly for water vapor \citep{Gordon17,ptashnik16}. Additionally, comparisons between atmospheric absorption data and HITRAN databases have been performed demonstrating overall excellent agreement, but identifying some regions with small differences between some HITRAN releases and observed line shapes, strengths, and locations (e.g. \citealt{toon16, example_hitran_mod}). Several atmospheric modeling codes for astronomical applications rely on HITRAN (e.g. \citealt{tapas,molecfit,telfit,terraspec}). Discrepancies at the 1-5\% level between observations and theoretical telluric models may be found when using older versions of the HITRAN database, although such discrepancies can also stem from the specific implementation of the radiative transfer calculation. In certain spectral regions, particularly in the region between the optical and near infrared, the on-sky and laboratory data and calculations underlying HITRAN database parameters may not be as robust as they are in the optical. It is therefore interesting to compare the results of our simplified telluric fits to the HITRAN database values. We do this only for water vapor due to the complexities and smaller number of lines in the case of molecular oxygen.

While our parameters are not accurate measures of the true underlying line parameters, we still expect to see trends between our line parameters and the physical quantities that describe how these lines vary with pressure and temperature. For example, in Figure \ref{fig:tel_shift} we show the difference between our best fit linecenters and the HITRAN catalog starting value plotted against the $\delta_{air}$ parameter that describes the magnitude of a pressure induced shift for a given line. Line transitions with a larger $\delta_{air}$ value will shift more at a specific pressure, which we observe. Since we fit multiple spectra simultaneously that were taken on different days and therefore under different atmospheric conditions, this induces extra scatter from averaging over different pressures and temperatures. However, since this trend largely depends on pressure, and higher in the atmosphere the pressure is consistently lower than the HITRAN reference pressure, the scatter induced by this fact does not wash out the overall trend. We note that we also see a correlation between the lower state energy level (elower) and the ratio of our optimized line strength to the HITRAN line strength values. However, this trend is slightly weaker due to other factors that determine how line strength changes. The same is true for the correlation between the line width ratio and $n_{air}$, the coefficient of temperature dependence on line broadening.

In making these comparisons, we observe a handful of outliers that are apparent in Figures \ref{fig:tel_shift} and \ref{fig:telbad_nu}. For the linecenters, we see one group shifted 0.07~cm$^{-1}$ down and another shifted 0.05~cm$^{-1}$ upward in frequency. Some of these correspond to lines that are also outliers when comparing $\gamma$ from our Lorentz fit to $\gamma_{air}$ from the HITRAN database (shown in Figure \ref{fig:telbad_nu}), as well as are discrepant in the line strength parameter, $S$. These lines are mostly located between 0.9-1.0~$\mu$m, which contains a strong water vapor band and has many saturated lines that often overlap making them difficult to fit and also introduces degeneracies in the best fit solution. Weak, unsaturated absorption features overlapping saturated regions would also have poorly constrained linecenters. An inspection of a subset of outlier points confirms that some of the outliers result from saturation issues. A subset can also be attributed to lines that border the edge of a fitting subregion and therefore are also poorly constrained. These outliers resulting from fitting-related causes show higher variance in their mean value determined from the 11 fits, as would be expected. However, another set of discrepant points exist that exhibit small variation in their line parameters between fits and under inspection are isolated or minimally blended with a neighboring line such that the most likely explanation for the discrepancy is the HITRAN catalog value itself. These lines are found across the entire spectral range analyzed here. We color the points in Figures \ref{fig:tel_shift} and \ref{fig:telbad_nu} by the most common references for the parameters $\delta_{air}$ and $\gamma_{air}$, respectively, but do not find that one source was the cause of the offsets, although the more recent works colored in orange in both plots (\citealt{JACQUEMART05} and \citealt{gamache04}) show less scatter. A more detailed study of these parameters could elucidate the observed discrepancies. For example, fitting the telluric output spectra from this work with a full atmospheric modeling code such as MOLECFIT \citep{molecfit} or TERRASPEC \citep{terraspec} would be a good framework for validating the results from this analysis.

\subsection{Discussion of the Telluric Model}
\begin{figure*}
\centering
\begin{minipage}{.45\textwidth}
  \centering
  \includegraphics[width=0.99\linewidth]{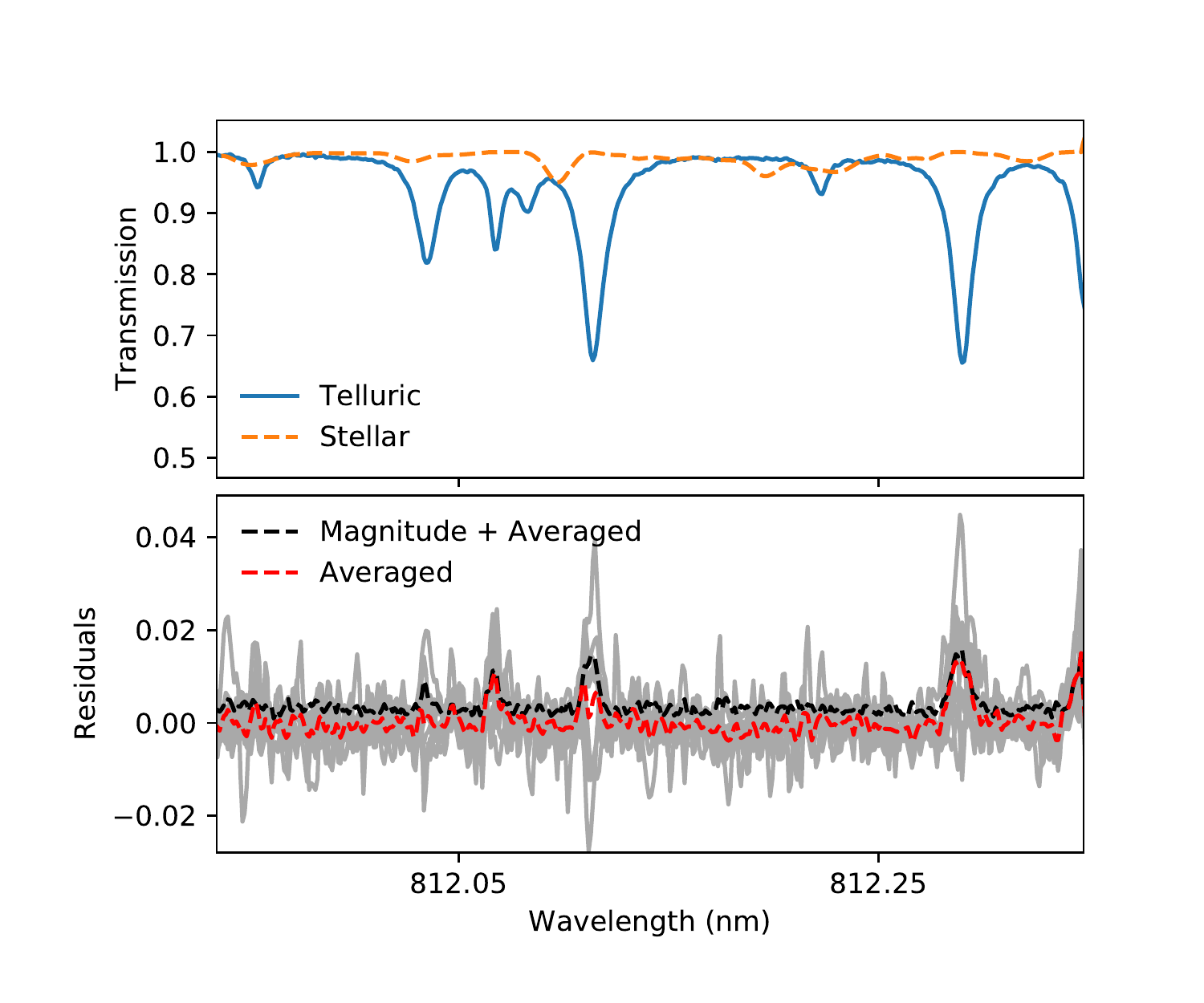}
    \caption{(Top) Stellar model in orange and median of extracted telluric spectra in blue from group 2 and (bottom) residuals from the fit in gray with the average shown in red and the average of their magnitudes shown in black. The residuals shown are from taking the telluric-corrected solar spectra and subtracting off the best fit solar spline model so they are centered at zero.}
    \label{fig:examp_resids}
\end{minipage}
\hfill
\begin{minipage}{.45\textwidth}
  \centering
      \includegraphics[width=0.99\linewidth]{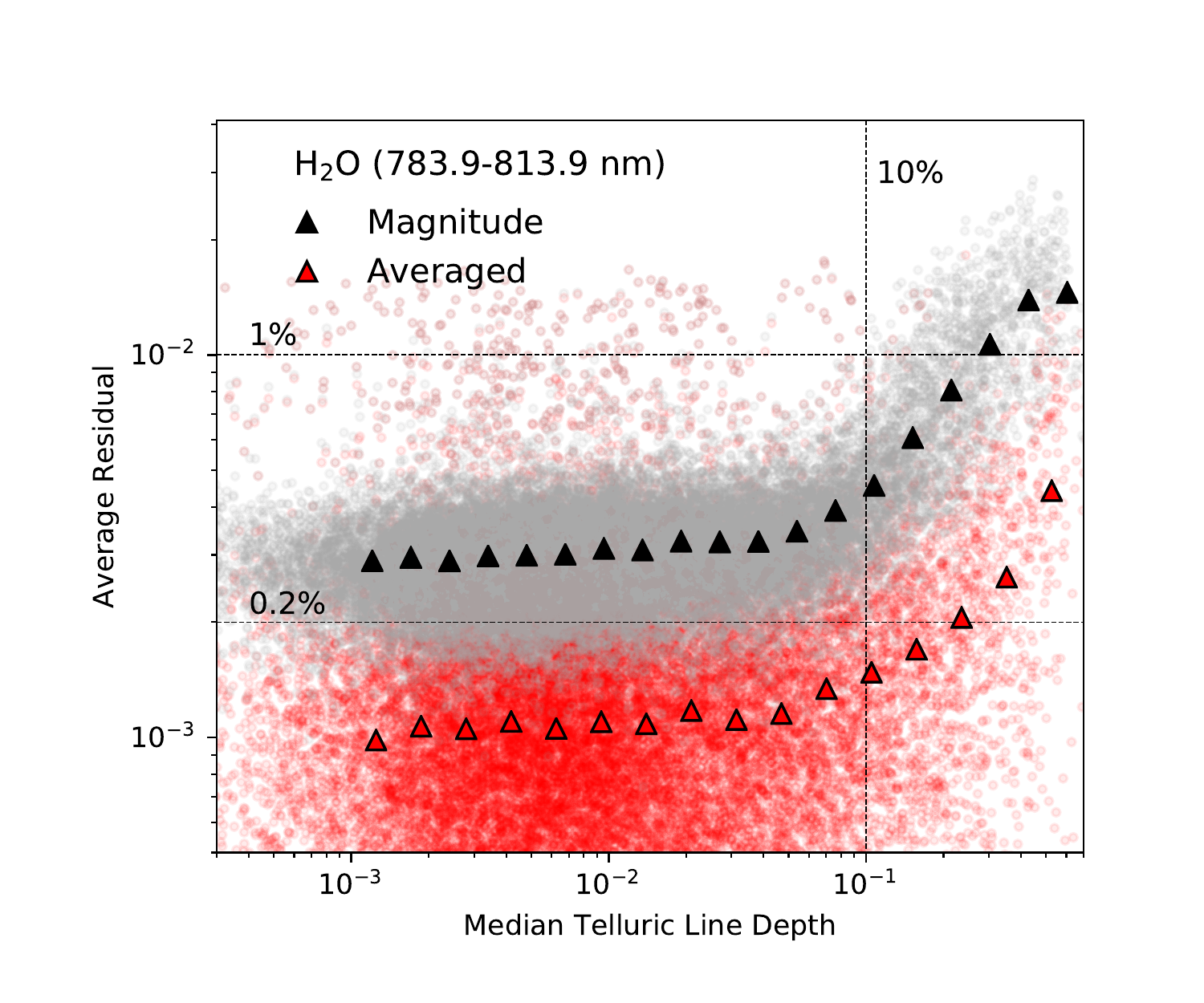}
    \caption{Water vapor average residual versus median telluric line depth for spectra in group 2. Here, a telluric depth of one corresponds to a saturated line. The residual array before averaging is defined to be the difference of each telluric-removed solar spectrum and the spline model. The gray points indicate values for which the absolute value was taken of the residual array before averaging, while for the red points the absolute value was taken after averaging. The corresponding triangles show the averages of the data in adjacent bins.}
    \label{fig:h2oresid}
  \end{minipage}
\end{figure*}

The routine used for fitting the telluric spectrum demonstrates the benefits of using a simple semi-empirical model for telluric fitting. Because both the spline and telluric models were analytic, this significantly sped up the fitting process and reduced the number of parameters defining our fit. A downside however is that the Lorentz profile is an approximation to the true underlying line shape that can also differ between observations due to the solar light passing through different lines of sight through the atmosphere that will have different pressure and molecular abundance profiles. Each absorption feature will change shape differently due to the nonuniform pressure and temperature dependencies of the transitions. Despite this simplification of our model, it still performs very well as can be seen in Figures \ref{fig:examp_resids} and \ref{fig:h2oresid}. Here, we show the residuals of our model against telluric line depth for a section of unsaturated water vapor features between 783.9-813.9~nm. In Figure \ref{fig:examp_resids} we show a subset of this region where we plot the median telluric spectrum on top and below we plot the residuals for group 2 data. We also show the magnitude of the residuals averaged for the 12 spectra in group 2 (black) that plotted as the gray points in Figure \ref{fig:h2oresid}. These demonstrate the typical residual value in  a single spectrum after dividing out the telluric lines. A second case is also shown where we allow the residuals to average down before taking the absolute value of the final array (red in both figures). This is characteristic of what happens when the solar atlas is generated (before replacing affected regions by the spline model) and we can see that the final remaining feature averages down better for some telluric lines depending on the residual structure. We can see that for both cases the magnitude of the residuals remains below 0.5\% for lines weaker than 10\% in depth with respect to the normalized continuum.

Most of the residuals in Figures \ref{fig:examp_resids} and \ref{fig:h2oresid} are due to not accounting for differences in the line shapes due to changing atmospheric conditions. A possible improvement could be to address this by parameterizing the atmospheric changes in time and modifying the telluric lines by utilizing the HITRAN parameters that describe these pressure and temperature line shape dependencies. Alternatively, a more empirical approach could be adopted, such as what was done in \cite{empirical_telfit_Leet19} or in the Wobble code developed by \cite{intro_bed19}. Wobble defines the telluric model by three principle components that are linearly combined; the measured flux in each spectral pixel value for each principle component spectrum is solved for directly. While Wobble would not work with solar data due to the small velocity shifts between the telluric and solar spectra, a physically motivated set of principle components could be used to fit the residuals from our model as was suggested by \cite{artigau14}, who also developed a principle component-based empirical telluric fitting algorithm. More investigation would need to be done to validate the usefulness of combining these two methods. 

Nevertheless, the telluric modeling code used in this work produces excellent results and the model may work well further into the NIR, where many Doppler precision spectrographs targeting K and M dwarfs are being operated in order to capture higher stellar line densities and fluxes. In particular, it avoids propagating any potential errors from line list databases that have been previously shown to affect atmospheric fits in the NIR \citep{bean_hitran_modify,hitran_errors}, and the ability to adapt the model to fit the radial velocity of stellar lines simultaneously could ultimately increase the fraction of a spectrum that can be used in the RV extraction process, that in the J band can be as high as 55\% of the region \citep{telluric_hband_loss}. With a larger barycentric velocity, this could be done for stellar targets without needing as extreme a range in airmass measurements as was required for the work presented here. The quick evaluation of this analytical model would also make up for the slow convolution step that would need to be added for fitting lower resolution data. 

\subsection{Micro-telluric Lines}
The impact of micro-telluric water vapor lines (lines having lower than $\sim$1\% depth relative to the continuum) is a growing concern to the field of high precision RV measurements that is pushing for the detection of terrestrial-sized exoplanets. Several studies have shown that micro-telluric lines, which are not visible after being convolved with an RV spectrograph's instrument profile, can skew RV measurements and be a large component of a survey's final error budget \citep{cunha14,sam_rv_err_budget,micro_telluric_peter,artigau14}. 

\begin{figure}
    \centering
    \includegraphics[width=0.9\linewidth]{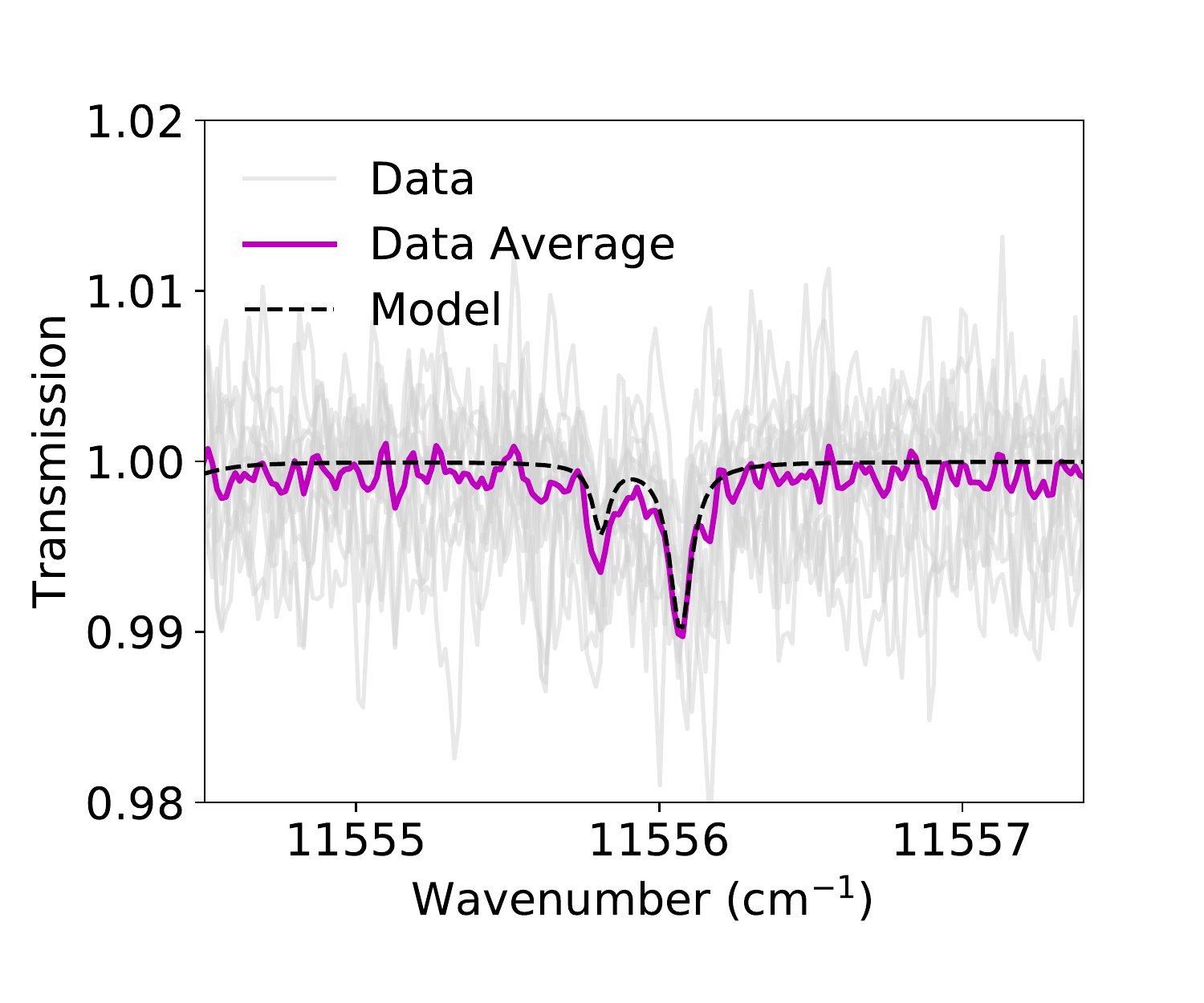}
    \caption{Demonstration of micro-telluric lines in the raw data. Here 13 telluric spectra are shown in gray with their average plotted in magenta and the telluric model in black. The two observed lines are an oxygen feature (left) and a water vapor feature (right) located in a NIR telluric window.}
    \label{fig:microtelluric}
\end{figure}

We point out that this telluric data set would be ideal for studying the temporal variations of micro-telluric line shapes since we are able to detect lines of depth 0.5\%-1\% compared to the continuum and binning multiple spectra in time or with similar airmass would help reduce the noise in the data to be able to study even weaker lines. We show a demonstration of two adjacent micro-telluric lines in Figure \ref{fig:microtelluric}, one due to molecular oxygen absorption and another due to water vapor absorption and show that these lines are clearly resolved in the average of 13 final telluric spectra. The water vapor line is weaker than our limit on lines included to be individually fit, however the uniform shift applied to these lines that was largely determined by the stronger features in the region did a good job aligning our model for the weaker telluric feature. Including these micro-telluric lines in the telluric model as we do here while solving for the radial velocity of the star may alleviate the impact they have on the radial velocity estimates. This should be confirmed in future work.

\section{Conclusions}\label{sec:conclude}
High resolution spectra of the Sun are important for many astrophysical studies including the study of stellar activity on Doppler spectroscopy, deriving the abundances of other stars, and understanding solar physics processes. High resolution spectra in which the individual solar lines are purely resolved are difficult to obtain from space and ground-based observations are plagued by telluric absorption features that move relative to the solar lines by a maximum of about a kilometer per second, which is not large enough for the stellar and telluric features to dissociate. Therefore, many of the stellar features overlapping telluric lines remain unreliable for analyses in high resolution solar spectra. Furthermore, the high signal-to-noise and high resolution of the telluric lines in FTS solar spectra also make them a useful dataset for studying micro-telluric lines, that are a poorly studies component in the error budget of next generation precision spectroscopy instruments. 

In this work we presented the telluric-corrected IAG solar flux atlas derived from observations taken in 2015 and 2016 in G{\"o}ttingen, Germany. We leverage the spread in airmass and Sun-Earth velocity to distinguish between spectral features that are either telluric or solar in origin and utilize a semi-empirical telluric model to separate the telluric lines from the solar data. We make available the final telluric-corrected solar spectrum online and additionally save the telluric spectra for possible use in studies such as investigating micro-telluric lines or validating various atmospheric models.

We find that our simplified telluric model works well with lines weaker than 10\% depth with respect to the continuum have residuals consistently below 1\% with their average being around 0.1\%. The addition of more molecular species would be possible for future work to extend this data reduction to the NIR portion of the IAG solar spectral data.

\section{Acknowledgements}
The authors would like to thank the anonymous referee for his or her comments that improved this manuscript. The authors also thank Dr. Iouli Gordon for his constructive comments on this work and the organizers of the 2019 Telluric Hack Week for hosting a nice week of talks and discussion that led to some of the methods incorporated into our final telluric model. This material is based upon work by ADB supported by the National Science Foundation Graduate Research Fellowship under Grant No. DGE-1321851.

\bibliography{solar_atlas_aas}   
\bibliographystyle{aasjournal}   

\appendix

\section{Justification of the Telluric Model} \label{sec:justify_tel}
In defining our telluric model we choose to fit each telluric line (in units of absorbance) as a simple Lorentz profile initiated with line parameters from the HITRAN database that are then individually optimized. The HITRAN line lists are extensively complete for O$_2$ and H$_2$O that we model here, making it a reliable starting point for our model. We chose to avoid radiative transfer modeling of Earth's atmosphere and instead use this semi-empirical approach in order to avoid propagating any errors in the line parameter database to our model with the added benefit of reduced computation times that enabled us to simultaneously fit multiple spectra at once. This allowed the model to leverage the different effects airmass and radial velocity have on the solar and telluric components of the spectra, which is crucial since the maximal range in solar RV ($\sim \pm$ 0.7~km s$^{-1}$) does not result in a separation of the two sources of spectral lines we ultimately wish to dissociate. Limiting the line shapes to a physically motivated functional form and placing these only where known atmospheric features exit is also useful to restrict the model to avoid over-fitting telluric residuals. In defining our telluric model this way, we are effectively simplifying the atmosphere to one constant temperature and pressure for all atmospheric heights.

\begin{figure}
    \centering
    \includegraphics[width=0.5\textwidth]{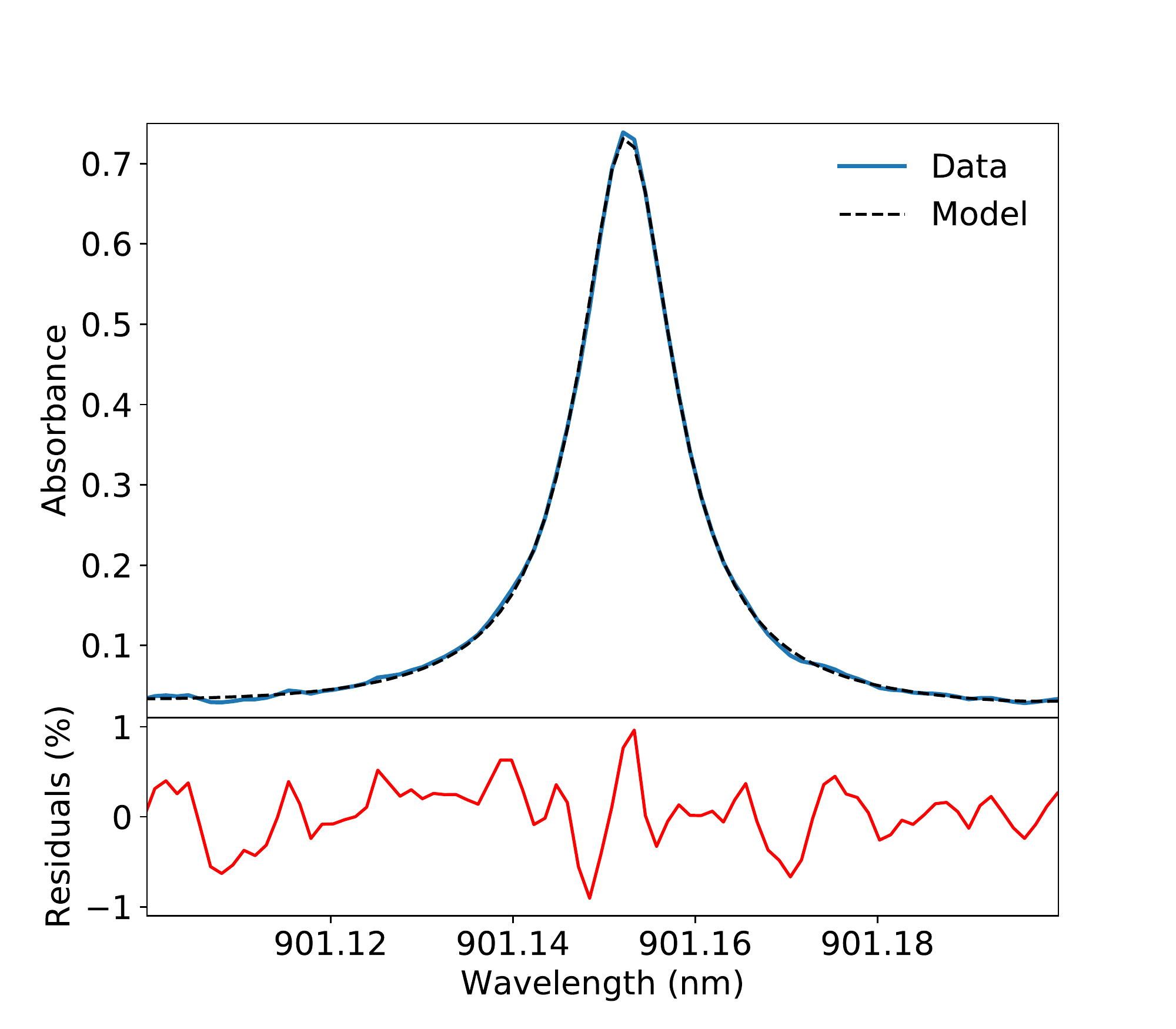}
    \caption{Example isolated water vapor feature along with a best-fit Lorentz profile.}
    \label{fig:lorentz}
\end{figure}

We find that this simplification and fitting each line individually as a Lorentz function is a good approximation and allows us to achieve residuals consistently below 2\% even for some saturated water vapor regions. We experimented with more complex line shapes, including Voigt, but found that for the water lines, the Lorenztian profile was a sufficient description of the line shape given our data\footnote{Although the water vapor line shapes are not necessarily symmetric and will change in time due to different atmospheric conditions along the line of sight, our fit optimizes an average line shape for the different observations and ultimately these residuals are reduced through averaging the spectra together}. \textbf{This can be seen in Figure \ref{fig:lorentz} where we show a nearly isolated water vapor transition with the best-fit solar spectrum removed. Comparing this strong line (50\% deep) to the best-fit Lorentz model produces residuals near the noise in the data with only an S-shaped structure nearing 1\% in magnitude in the core of the line.} This was expected since the water vapor lines are formed at lower altitudes and therefore are primarily pressure-broadened \citep{Hedges16}. This is not necessarily true of oxygen since it extends into the upper atmosphere and therefore has a larger Doppler broadening component making a Voigt profile a better approximation. The resulting s-shaped residuals from the simplification of the line shapes for both species vary at a frequency that is typically too high for our stellar spline model to capture (see \S \ref{sec:spline}) and do not correlate with the solar radial velocity. Therefore, we can simply replace regions overlapping dense telluric absorption with the spline fit to remove the majority of these telluric residuals. Because we perform our fits on 11 groupings of data, any telluric residuals captured by our spline model should vary in strength between the 11 best-fit stellar models and can therefore still be identified as unreliable.

\section{Basis Splines}\label{sec:bsplines}

The basis splines, $B_{j,q;t}$, are calculated internally by the \texttt{BSpline} function and are defined recursively as follows using $t_j$ as the knot locations that match the units of $\nu$:

\begin{equation}
    B_{j,0}(\nu) = 1, \:\mathrm{if}\: t_j \leq \nu \leq t_{j+1}\:, \mathrm{otherwise}\: 0,
\end{equation}

\begin{equation}
    B_{j,q}(\nu) = \frac{\nu - t_j}{t_{j+q} - t_j} B_{j,q-1}(\nu) + \frac{t_{j+q+1}  -\nu}{t_{j+q+1} - t_{j+1}} B_{j-1,q-1}(\nu) .
\end{equation}

\section{Zero-point Offset from Telluric Lines}\label{sec:zpo}
We also investigated using the oxygen lines for the purpose of finding an accurate wavelength scale but found that uncertainties in the fitted line centers and the $\delta_{air}$ parameters that determine each line's pressure-induced shift made it difficult to calculate a more accurate estimate of a zero offset velocity than can be achieved using the iodine template. Fitting a line to the line-center shifts found in this work versus the HITRAN $\delta_{air}$ parameters allows us to solve for a zero offset velocity that would shift the fitted line to cross the origin (i.e. an extrapolated $\delta_{air}$ value of zero induces no measured line shift). Doing this, we found the offset to be 0.005~cm$^{-1}$+/-0.004~cm$^{-1}$ using all oxygen lines for which our measured line centers had uncertainties of less than 6$\times 10^{-4}$ as determined by the standard deviation of the values found in our 11 fits. For $\delta_{air}$ we assumed a uniform uncertainty of 5$\times 10^{-5}$, which reflects what is reported in \cite{o2_pres_shift}. We note that the evaluated zero point offset depends on the source of the measurement and the isotopologue number, which is likely due to offsets in $\delta_{air}$, which are calculated in reference to the lines of another species, typically either iodine or carbon monoxide, and which species is chosen will offset the measured value. While the water vapor lines have higher $\delta_{air}$ uncertainties \citep{water_shift}, there are also more data points and together they provide estimate of 0.0063~cm$^{-1}$+/-0.0020~cm$^{-1}$. Both of these estimates are slightly sensitive to the errors assumed for $\delta_{air}$, but remain consistent with the zero offset velocity determined using the iodine lines.

\end{document}